\documentclass[english,manuscript, superscriptaddress]{revtex4}
\usepackage{times}
\usepackage[T1]{fontenc}
\usepackage[latin1]{inputenc}
\usepackage{graphicx}

\makeatletter

\usepackage{babel}
\makeatother
\begin{document}

\title{Charged defects in graphene and the ionicity of hexagonal boron nitride
in direct images}

\author{Jannik C. Meyer}

\affiliation{Electron microscopy group of the materials sciences, University of
Ulm, 89081 Ulm, Germany}

\author{Simon Kurasch}

\affiliation{Electron microscopy group of the materials sciences, University of
Ulm, 89081 Ulm, Germany}

\author{Hye Jin Park}

\affiliation{Max Planck Institute for solid state research, Heisenbergstr. 1,
70569 Stuttgart, Germany}

\author{Viera Skakalova}

\affiliation{Max Planck Institute for solid state research, Heisenbergstr. 1,
70569 Stuttgart, Germany}

\author{Daniela Künzel}

\affiliation{Institute of Theoretical Chemistry, University of Ulm, 89069 Ulm,
Germany}

\author{Axel Groß}

\affiliation{Institute of Theoretical Chemistry, University of Ulm, 89069 Ulm,
Germany}

\author{Andrey Chuvilin}

\affiliation{Electron microscopy group of the materials sciences, University of
Ulm, 89081 Ulm, Germany}

\affiliation{CIC nanoGUNE Consolider, San Sebastian, and Ikerbasque, Basque Foundation
for Science, Bilbao, Spain}

\author{Gerardo Algara-Siller}

\affiliation{Electron microscopy group of the materials sciences, University of
Ulm, 89081 Ulm, Germany}

\affiliation{Technical University Ilmenau, Ilmenau, Germany}

\author{Siegmar Roth }

\affiliation{Max Planck Institute for solid state research, Heisenbergstr. 1,
70569 Stuttgart, Germany}

\affiliation{School of Electrical Engineering, WCU Flexible Nanosystems,\\
Korea University, Seoul, Korea}

\author{Takayuki Iwasaki}

\affiliation{Max Planck Institute for solid state research, Heisenbergstr. 1,
70569 Stuttgart, Germany}

\author{Ulrich Starke}

\affiliation{Max Planck Institute for solid state research, Heisenbergstr. 1,
70569 Stuttgart, Germany}

\author{Jürgen Smet}

\affiliation{Max Planck Institute for solid state research, Heisenbergstr. 1,
70569 Stuttgart, Germany}

\author{Ute Kaiser}

\affiliation{Electron microscopy group of the materials sciences, University of
Ulm, 89081 Ulm, Germany}

\begin{abstract}
We report on the detection and charge distribution analysis for nitrogen
substitutional dopants in single layer graphene membranes by aberration-corrected
high-resolution transmission electron microscopy (HRTEM). Further,
we show that the ionicity of single-layer hexagonal boron nitride
can be confirmed from direct images. For the first time, we demonstrate
by a combination of HRTEM experiments and first-principles electronic
structure calculations that adjustments to the atomic potentials due
to chemical bonding can be discerned in HRTEM images. Our experiments
open a way to investigate electronic configurations in point defects
or other non-periodic arrangements or nanoscale objects that can not
be studied by an electron or x-ray diffraction experiment.
\end{abstract}
\maketitle

\section{Introduction}

The elastic scattering of high-energy electrons within a material,
as detected by electron diffraction, electron holography or high resolution
transmission electron microscopy (HRTEM), probes the distribution
of the internal electrostatic potential \cite{KirklandEMcomputing,Buseck_HRTEMandAssoc}
of a sample. This potential is given by the Coulomb potential of the
nuclei, screened by the electrons. The dominant contribution to the
electrostatic potential comes from the core region, i.e. the region
close to the nucleus. However, only the valence electrons participate
in chemical bonds and are rearranged depending on the atomic configuration.
Therefore, HRTEM images are usually dominated by structural information,
i.e. the position of the atoms rather than the electronic configuration.
The effect of bonding on HRTEM image contrast has been explored in
previous studies \cite{AbInitioHRTEMGeming98,AbInitioInterfaceTEMMogck04,DengMarksBonding06,DengGlowingDefects07};
however, it was concluded that the predicted effects would be difficult
to detect experimentally. This has led to the assumption that conventional
(independent atom model, IAM) HRTEM image simulation is sufficient
for structural and elemental identification. The effect of binding
electrons has been detected experimentally in electron diffraction
experiments \cite{SpenceBondMap99,CBEMoribe02}, but so far was not
discerned in TEM images. However, diffraction experiments are limited
to periodic structures and sufficiently large samples. Here, we study
these chemical shifts for a point defect, the nitrogen dopant \cite{NdopedGr09}
in graphene \cite{Novoselov2004Sci,GeimGrapheneReview07}, and few-nm
sized hexagonal boron nitride mono-layer membranes \cite{MeyerBNnl09,IijimaBN09,ZettlHbn2009,KrivanekBNnature10}.
The precisely defined sample geometries in terms of thickness (one
layer), amorphous adsorbates and defects (none in the selected regions),
along with very high sample stability and therefore unprecedented
signal to noise ratios (by using lower acceleration voltages to avoid
sample damage) enable sufficiently accurate measurements \cite{DengMarksBonding06}.
Both, HRTEM experiments and first-principles electronic structure
calculations, show that the IAM is not sufficient for accurate TEM
image simulations of these materials. Our experiments open a new way
to study electronic configurations, in particular for point defects,
other non-periodic arrangements, or nanoscale objects that can not
be analyzed in a diffraction experiment.

\section{Methods summary}

We prepared nitrogen doped graphene membranes by following the CVD
methods for graphene synthesis \cite{CVDKimNature09,CVDCopperRuoff09,CVDNiHyeJin10},
with the addition of small amounts of ammonia as nitrogen source \cite{NdopedGr09}
during the growth. We have used both the CVD growth on nickel substrates
and on copper substrates, and in both cases the addition of ammonia
into the reaction has led to nitrogen doped graphene sheets. We transfer
the CVD grown graphene sheets to commercial TEM grids as described
previously \cite{CVDNiHyeJin10}. Single-layer hexagonal boron nitride
is prepared and imaged as described previously \cite{MeyerBNnl09}.
TEM imaging is carried out using an image-side aberration corrected
Titan 80-300, operated at 80kV. The spherical aberration is set to
ca. 20\ensuremath{µ}m and a defocus of $f_{1}=-90\textrm{\AA}$ (Scherzer
defocus) and $f_{2}=-180\textrm{\AA}$ (Graphene lattice in the second
extremum of the contrast transfer function (CTF)) is used. Drift-compensated
averages of 20-40 CCD exposures are used. Post-filtering of the images
is needed to discover the N substitution defects in our experimental
data. Shown here is the example of a Fourier filter that removes the
periodic components of the graphene lattice, and a simple gaussian
low-pass filter chosen so that the graphene lattice contrast is reduced
to ca. 0.5\% (in this case the N substitution and graphene lattice
appear with similar contrast). TEM image simulations for the bonded
atomic configuration follows the procedure pioneered by Deng and Marks
\cite{DengMarksBonding06} with some modifications. In brief, a relaxed
atomic configuration is obtained using the VASP DFT code \cite{Kresse96}
which is well-suited to perform large scale geometry optimizations
in graphene-like systems from first principles \cite{Kuenzel2009}.
Then, the WIEN2k DFT code \cite{WIEN2k2001} is used to obtain the
all-electron self-consistent electron density and corresponding electrostatic
potentials for this configuration. The TEM image simulation is then
based on projections of this electrostatic potential, and follows
the approximations for thin specimen as in Chapter 3 of Ref. \cite{KirklandEMcomputing}.
Further details of the sample preparation and characterization, DFT
calculations, potential calculations, TEM image simulation, further
TEM images, and a detailed analysis of residual aberrations are described
in the supplementary information.

\section{Nitrogen doped graphene}

\begin{figure}
\includegraphics[width=0.56\linewidth]{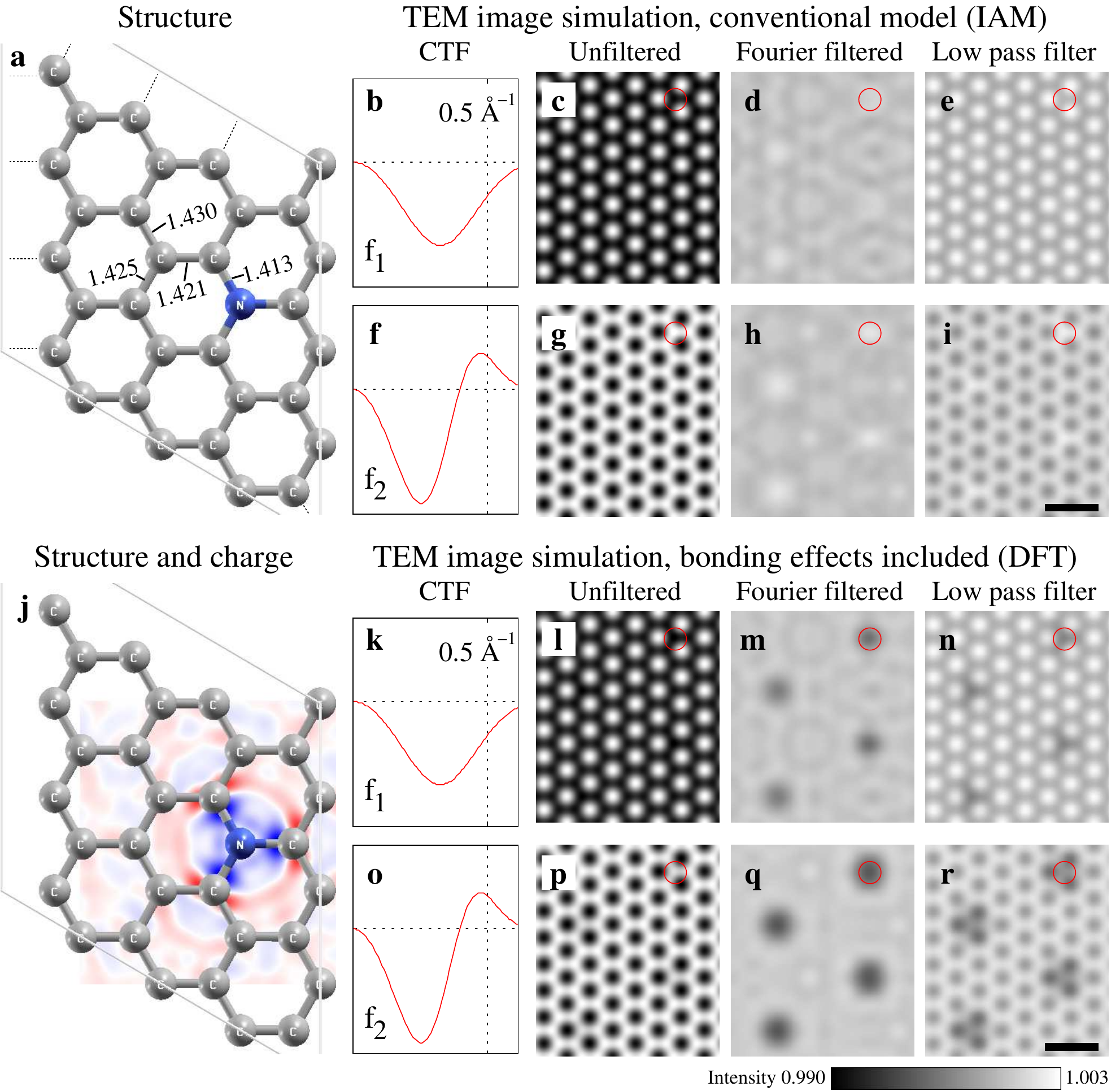}

\caption{TEM simulations for nitrogen doped graphene. (a) Relaxed atomic configuration
for a nitrogen substitution in graphene. Bond lengths are given in
$\textrm{\AA}$. (b-i) Conventional, independent-atom model (IAM)
based TEM image simulation for different defoci and filters (see text).
One of the four nitrogen substitutions is marked with a red circle.
(j) Atomic structure (same bond lenghts), with the differences in
projected electron density between DFT and IAM model shown as colors
(periodic components removed -- see Fig. \ref{fig:ExpAndSim}f). Blue
corresponds to a lower, red to a higher electron density in the DFT
result as compared to the neutral-atom IAM case. (k-r) TEM image simulations
based on the electrostatic potentials of the all-electron DFT calculation.
The greyscale calibration bar applies to (d,e,h,i,m,n,q,r), which
are all displayed on the same greyscale range for direct comparison.
Panels b,f,k,o show calculated contrast transfer functions (CTF) for
the respective row of images. The scale bars are $5\textrm{\AA}$.\label{fig:Nsim} }
\end{figure}

Nitrogen doped graphene in our case refers to a substitutional doping,
i.e. a small fraction of carbon atoms (ca. 0.1\%) was replaced by
nitrogen during synthesis. This configuration has been considered
in many previous works (e.g. \cite{NsubstDistAndCH05,NsubstCNTbondlength07}),
and it was found that the C-N bond length in this configuration is
nearly identical to the C-C bond length in graphene (with differences
less than 2pm). Our own DFT-relaxed configuration (Fig. \ref{fig:Nsim}a)
is in agreement with these earlier works, and is used for all TEM
simulations of this defect.

Fig. \ref{fig:Nsim} shows calculated TEM images with and without
the inclusion of bonding effects. Images labeled as {}``IAM'' (Fig.
\ref{fig:Nsim}b-i) are based on the so-called independent atom model.
In this model, the potential and charge distribution of the specimen
is calculated as a superposition of potentials and charge densities
that have once been calculated for an isolated atom of each element.
This is the conventional model for TEM image simulation. We show here
two different imaging conditions (defocus $f_{1}$ and $f_{2}$),
and two filters (which are needed to discover the defect in experimental
data), as described in the methods. Remarkably, the nitrogen substitution
defect is expected to be practically invisible (less than 0.1\% contrast)
within the IAM, i.e., when bonding effects are not taken into account.
On the contrary, charge densities and potentials obtained from the
DFT calculation naturally contain the electronic configuration of
the bonded situation (corresponding images are labeled with {}``DFT'').
TEM image simulations based on the accurate electrostatic potential
from the DFT calculation are shown in Fig. \ref{fig:Nsim}k-r. Here,
we expect a detectable signal when using the DFT based potentials
for TEM image simulation: The N substitution is now visible as a weak
dark spot, best seen in the filtered images. In the comparison, there
is even a change of sign, i.e, a (very weak) white spot in the Fourier
filtered images of the IAM (Fig. \ref{fig:Nsim}d,h) while there is
a stronger (and detectable) dark spot according to the DFT model (Fig.
\ref{fig:Nsim}m,q).

There is a remarkable focus dependence in the DFT based image contrast,
shown here by the two different defocus values $f_{1}$ and $f_{2}$.
Defocus value $f_{1}$ represents a standard condition where spherical
aberration and defocus are tuned to obtain a single pass-band in the
contrast transfer function (CTF, Fig. \ref{fig:Nsim}b) that ends
approximately at the information limit \cite{LentzenReview06}. In
this case, shown in Fig. \ref{fig:Nsim}b-e, the nitrogen substitution
is expected to produce a weak dark spot (0.3\% contrast); which is
still difficult to detect since it is both a weak and narrow signal.
Using a larger defocus ($f_{2}$), however, both width and amplitude
of the signal increases (Fig. \ref{fig:Nsim}f-i). This focus behaviour
is contrary to what would be expected for a point scatterer, such
as an ad-atom or a charge that is localized on one atom. For this
reason, detection of the nitrogen substitution defect in graphene
is easier (requiring significantly lower electron doses per area)
when using the larger defocus settings. 

\begin{figure}
\includegraphics[width=1\linewidth]{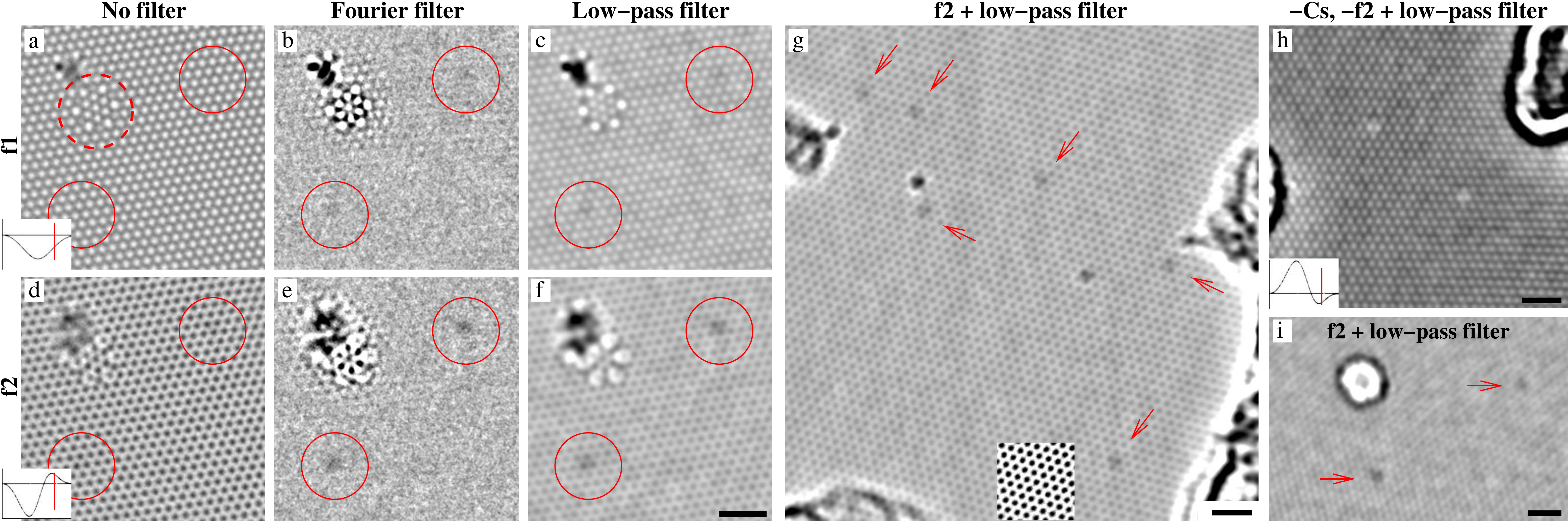}

\caption{Nitrogen dopants in graphene. Imaging conditions and filters in (a-f)
correspond to Fig. \ref{fig:Nsim}. In the Scherzer defocus image
(a-c), dark contrast can be directly interpreted as atomic structure.
However, the nitrogen substitution defects are not significantly above
the noise (red circles, a-c). In a larger defocus image ($f_{2}$,
d-f) with similar noise level, the nitrogen defects are clearly detected
as a smooth dark contrast (in any case, a filter (e,f) is needed to
discern the N dopants against the much stronger signal of the single
layer graphene lattice). The extended defect (red dashed line in (a))
allows to compare the same atomic position in both focus values. (g)
Image from a larger area, showing six nitrogen substitution defects
marked by red arrows. (h) Image of two nitrogen substitutions obtained
with a reversed CTF (negative spherical aberration $C_{s}$, positive
defocus), showing the substitution positions as white areas. (i) Irradiation-induced
nitrogen substitutions (arrows, see text). The strong feature on the
upper left is a beam-induced hole. Red lines in the CTF plots (insets)
indicate the $2.13\textrm{\AA}$ graphene lattice spacing. All scale
bars are 1nm. \label{fig:N1}}
\end{figure}

Experimentally, we identify single-layer graphene areas in the nitrogen
doped, CVD grown graphene sheets by an electron diffraction analysis
as described previously \cite{MeyerGrapheneNature07,MeyerGrapheneSSC07}.
We record long image sequences, >30 exposures with a pixel size of
$0.2\textrm{\AA}$ and ca. $10^{4}$ counts per pixel, for each value
of defocus. Importantly, the graphene structure and defects of interest
remain stable throughout the long exposure and associated high dose.
Then, a drift compensated average of 39 (Fig. \ref{fig:N1}a-c) or
34 (Fig. \ref{fig:N1}d-f) exposures provides the very high signal
to noise ratio images that are further analyzed. As predicted by the
DFT model, they exhibit a weak dark contrast in the larger defocus
images (Fig. \ref{fig:N1}d-f), but disappear in the noise as the
focus is set to the optimum (Scherzer defocus) conditions (Fig. \ref{fig:N1}a-c).
Fig. \ref{fig:N1}a-f are from a nitrogen doped sample synthesized
on a nickel surface. Fig. \ref{fig:N1}g shows an example from N doped
graphene grown on copper. Fig. \ref{fig:N1}h shows two nitrogen substitutions
imaged with reversed contrast transfer (negative Cs, positive defocus),
where the substitution atoms are revealed as weak white spots. Finally,
Fig. \ref{fig:N1}i shows an alternative way to fabricate atomic substitutions
in graphene: Here, an un-doped graphene sample was briefly exposed
to higher energy electrons (300kV, ca. $10^{7}\frac{e^{-}}{\textrm{nm}^{2}}$),
and subsequently imaged at 80kV to prevent further damage. After such
a treatment, we also find nitrogen substitution defects. A likely
mechanism is a substitution of beam-generated vacancies by atoms from
the residual gas. However, we note that also a variety of other defects
is generated by this approach.

The focus dependence of the nitrogen defect in graphene deserves further
discussion. As noted above, a contrast that becomes both wider and
stronger with larger defocus is contrary to expectations for a point
scatterer, such as an ad-atom or a localized charge. It means that
the experimentally accessible information, after removing the periodic
component of the graphene lattice, contains predominantly lower spatial
frequency information. This information is cut off by the CTF (Fig.
\ref{fig:Nsim}b,k) in the optimum-focus ($f_{1}$) image of the aberration-corrected
microscope. At the larger defocus, the CTF (Fig. \ref{fig:Nsim}f,
o) shows oscillations but includes more of the lower spatial frequencies,
thus revealing the charge associated with the nitrogen defect. It
is important to note that we can draw this conclusion already from
the experimental data, rather than only from the calculated charge
distribution (as also shown below): The focus dependence confirms
from experimental data, that the signal we detect here is not localized
on the substitution atom, but spread onto a larger area. 

\begin{figure}
\includegraphics[width=0.8\linewidth]{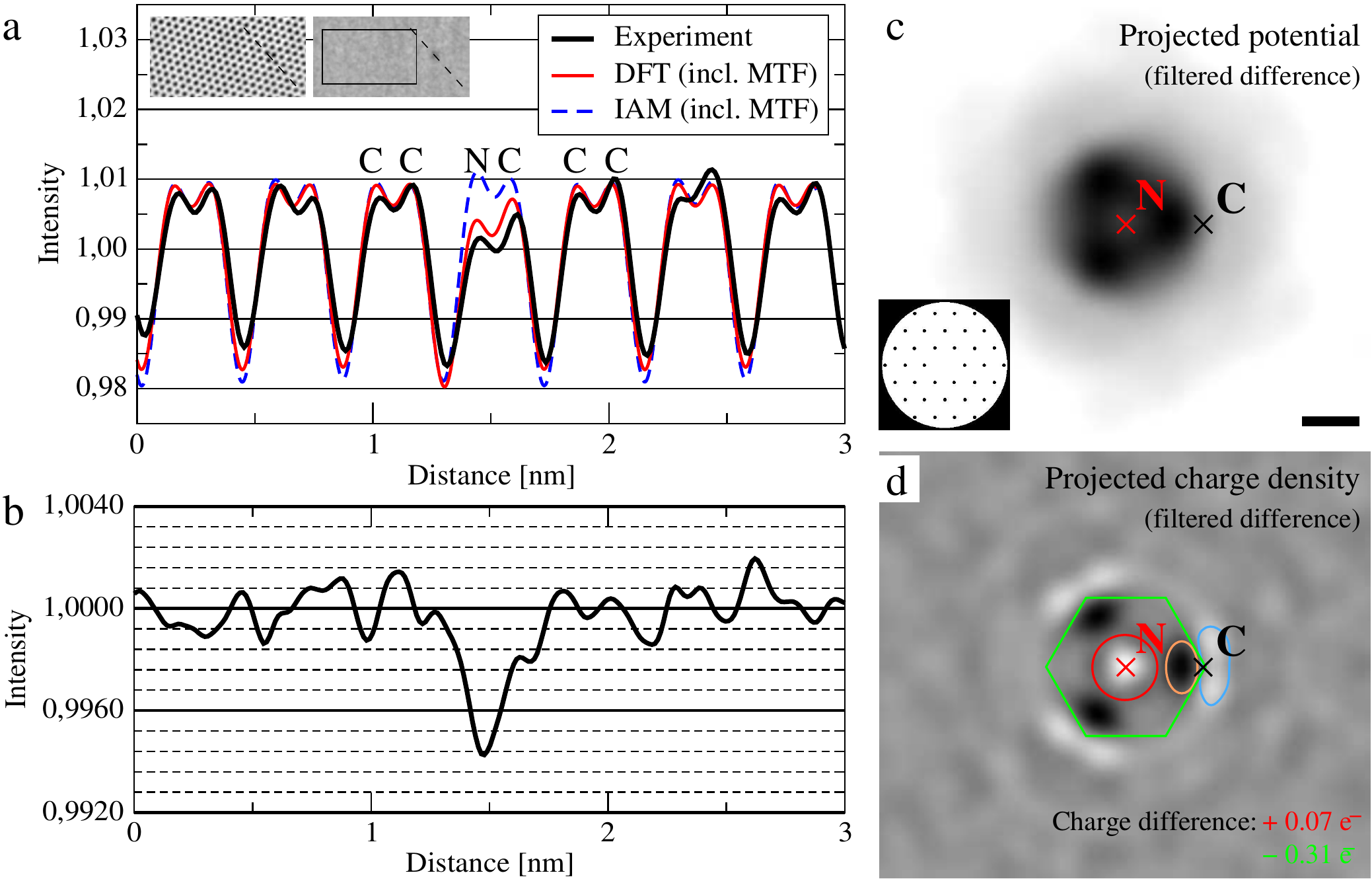}

\caption{Analysis of the nitrogen defect. (a) Comparison between experiment
and simulations based on IAM and DFT potentials for defocus value
$f_{2}$. Inset shows the image and profile, and the Fourier filtered
image (graphene lattice removed). The noise level is measured as the
standard deviation in a featureless region of the filtered image,
close to the defect (black box). A standard deviation of 0.0008 is
found in this example. (b) Line profile with the periodic components
removed. The intensity dip at the nitrogen defect is 7x stronger than
the standard deviation, and thus clearly above the noise. (c-d) Analysis
of differences between independent-atom and DFT based electronic configuration.
(c) Difference in the projected potentials (DFT-IAM), with periodic
components of the graphene lattice removed. The filter is indicated
as inset. Higher projected potentials are shown in dark contrast,
in agreement with our TEM imaging conditions. (d) Difference in projected
charge density (DFT-IAM), with the same filter. The dipole-shaped
charge on the neighbouring carbon atoms is highlighted in one example
by orange and blue lines. Total charge difference in the red cylinder
(N atom) and green hexagon shaped area (region up to neighbouring
carbons) is indicated. Scale bar for c+d is $1\textrm{\AA}$. \label{fig:ExpAndSim}}
\end{figure}

A quantitative comparison between the simulation and experiment is
shown in Fig. \ref{fig:ExpAndSim}. For this comparison, the modulation
transfer function (MTF) of the CCD camera was measured and applied
to the simulated images, following the procedure of Ref. \cite{ThustMTF09}.
In addition, a small ($0.7\textrm{\AA}$ FWHM) gaussian blur was applied
to both, unfiltered experimental data and simulation, in order to
reduce the pixel noise in the experimental data. As can be seen in
Fig. \ref{fig:ExpAndSim}a, a good match of the experimental profile
and the DFT based simulation is found. We then remove the periodic
component of the graphene lattice, in order to estimate the noise
level in this image. We obtain a standard deviation of 0.0008 in this
case and a signal of the nitrogen substitution that is 7x above this
noise (Fig. \ref{fig:ExpAndSim}b). 

While the TEM image depends directly on the electrostatic potentials,
it is useful to look at both, the projected potentials and charge
density of the calculations, in order to understand how the charges
are re-arranged in the bonded configuration. Fig. \ref{fig:ExpAndSim}c
shows the difference in projected potentials, with the periodic components
of the graphene removed (the unfiltered images are given in the supplement).
Importantly, there is a \emph{spatially extended} dark signal, with
a diameter of ca. $2.4\,\textrm{\AA}$ (twice the C-N bond length)
in the projected potential. Fig. \ref{fig:ExpAndSim}d shows the difference
in projected charge densities, again with the periodic components
removed. Here, one can see that the most dominant effect in the charge
distribution is not on the nitrogen atom itself, but the dipole-shaped
rearrangement of the electrons on the nearest-neighbour carbons around
the nitrogen defect. 

As a side remark, we note that the nitrogen defect is reactive \cite{NsubstDistAndCH05},
in comparison to the inert carbon surface. In some cases, we observe
a repeated trapping and de-trapping of contamination atoms on the
nitrogen defect in long image sequences (see supplementary information).
The ad-atoms on the nitrogen defects are surprisingly stable, such
that a few atomically resolved images can be obtained before the configuration
changes under the beam. The reactivity of the nitrogen substitution
might therefore be useful for the potential application of graphene
as a TEM substrate, as it allows to trap objects while only minimally
affecting its transparency.

\section{Results and Discussion: Hexagonal Boron Nitride}

\begin{figure}
\includegraphics[width=0.5\linewidth]{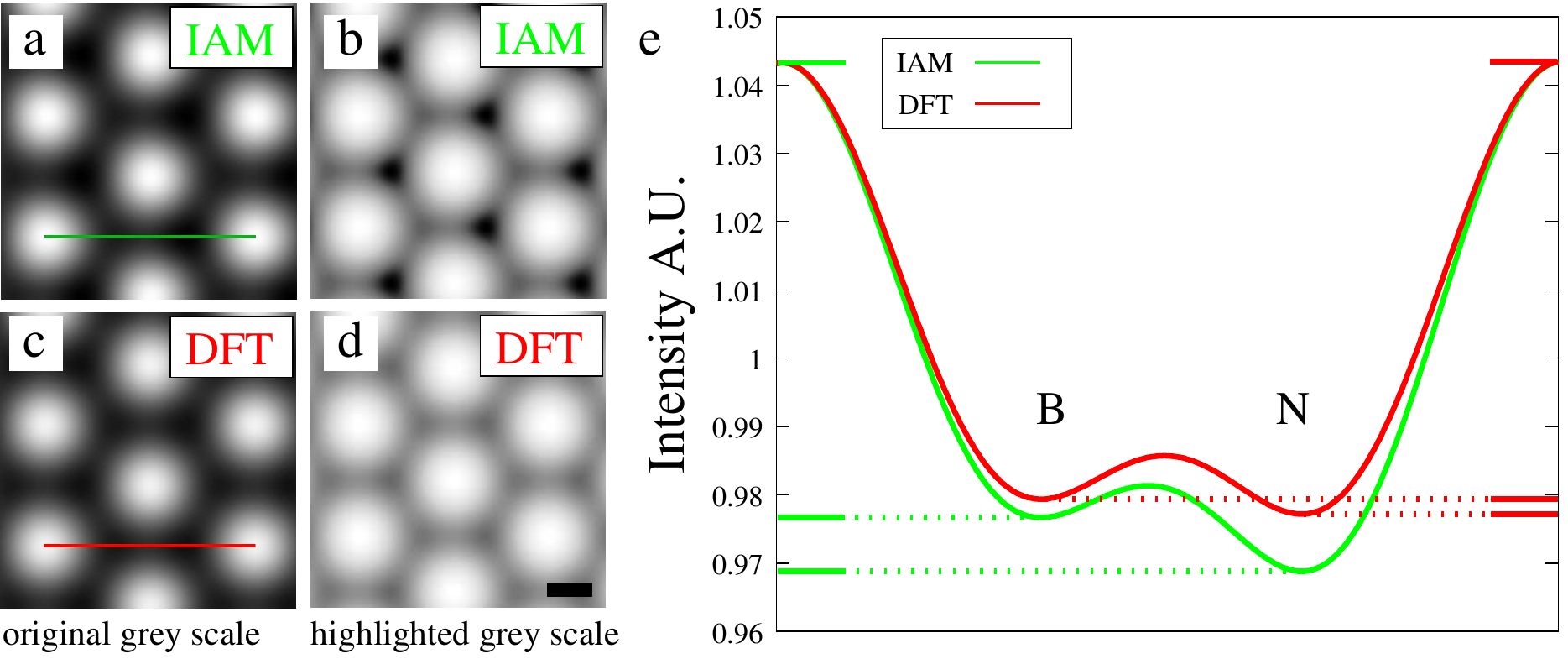}

\caption{Simulated TEM images for hexagonal boron nitride. (a+b) Conventional,
independent atom model (IAM) TEM image simulation for single-layer
hexagonal boron nitride. (c+d) TEM simulation using potentials from
the all-electron DFT simulation. (e) Intensity profile plots for the
two simulations. Scale bar in (d) is 1$\textrm{\AA}$. \label{fig:BNsim}}
\end{figure}

We now turn to the calculations and measurements for the single-layer
hexagonal boron nitride membranes. Hexagonal boron nitride (hBN) is
a periodic structure with well known partially-ionic character \cite{PaulingBNandGrPNAS1966}.
However, single-layer hBN membranes as obtained in a TEM are only
a few nanometer in size \cite{MeyerBNnl09,IijimaBN09,ZettlHbn2009,KrivanekBNnature10}
and therefore too small for an analysis by electron or x-ray diffraction.
By using the direct image, we can choose a nanometer-sized area, which
is in addition free of any defects. 

The conventional independent atom model (IAM) TEM simulation is shown
in Fig. \ref{fig:BNsim}a+b, while Fig. \ref{fig:BNsim}c+d shows
the TEM image simulation for the DFT based electrostatic potentials
for the bonded configuration. The difference between the IAM and DFT
model has important implications: The IAM based simulation predicts
a significant contrast difference for the two elements, already with
only $2.16\,\textrm{\AA}$ (1-100 reflection) resolution. Thus, based
on the IAM one would expect a different contrast for the B and N sites
as soon as lattice resolution is achieved. The DFT based simulation,
on the other hand, predicts a practically symmetric image in this
case. It has to be considered as a coincidence that the elemental
discrimination is not possible (at this resolution) when using the
DFT based TEM simulation. We can interpret the DFT result in such
a way that, due to the ionicity of h-BN, the contrast difference that
is expected for neutral atoms is almost exactly canceled: The accumulation
of electron charge on the N site, already noted by Pauling in Ref.
\cite{PaulingBNandGrPNAS1966}, leads to a stronger screening of its
core potential and thus a reduced contrast for this element (Fig.
\ref{fig:BNsim}e, see also supplementary info). Thus, we can confirm
the DFT prediction and verify the ionicity of single-layer hBN from
HRTEM images, even though we can not assign the B and N sublattices.

\begin{figure}
\includegraphics[width=0.72\linewidth]{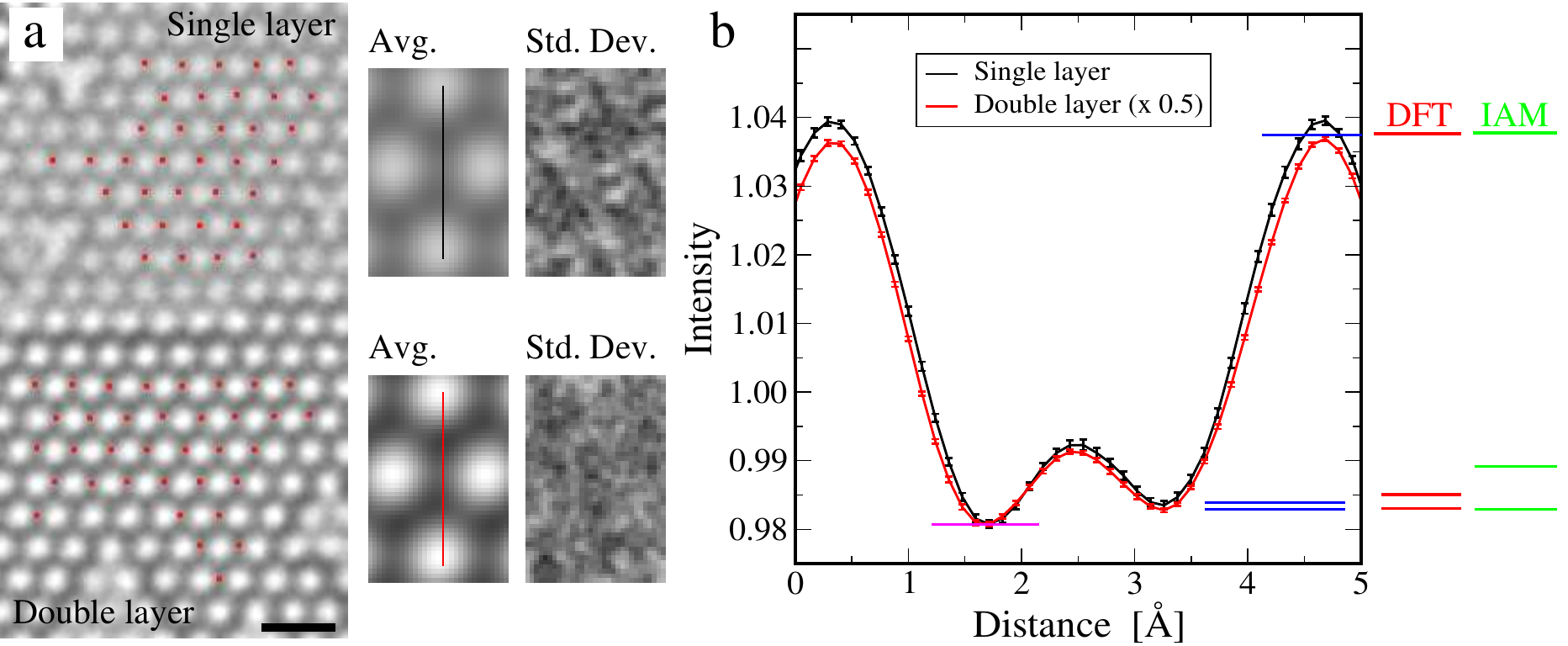}

\caption{Experimental data for hexagonal boron nitride. (a) Image of single-
and bi-layer hBN (scale bar is $5\textrm{\AA}$). The unit cells indicated
by the red dots were chosen for analysis (away from defects and edges),
panels on the right hand side show the average and the (featureless)
standard deviation. (b) Intensity profiles from single- and double
layer average (standard deviation as error bars). The contrast of
the double layer was numerically reduced by a factor of 2 for comparison.
Intensity minimum on the left sublattice was shifted to the same value
(pink line), and then the intensity on the right sub-lattice is compared
(blue lines). Contrast difference as expected from IAM and DFT model
is indicated.\label{fig:BNexp}}
\end{figure}

For the experimental case, the separation of intrinsic contrast (i.e.,
that of the sample) and effects of optical aberrations (from the microscope)
is difficult and technical, and therefore discussed in more detail
in the supplementary information. At this point, we only show the
comparison between single-layer and bi-layer hBN. For a bi-layer,
a symmetric profile is expected due to the symmetry of the projected
structure (B is above N). Hence, any asymmetry in the bi-layer profile
represents residual electron optical aberrations only. Assuming identical
imaging conditions for all points of the sample, the comparison between
a single-layer and bi-layer region would show the {}``intrinsic''
contrast difference in the mono-layer, if present. The experimental
result is shown in Fig. \ref{fig:BNexp}: Fig. \ref{fig:BNexp}a shows
the direct image (average of 21 exposures), with 35 selected unit
cells indicated in both areas, and their average and standard deviation.
Fig. \ref{fig:BNexp}b shows the corresponding intensity profiles.
Shown here as error bar is the standard deviation as mean deviation
of individual unit cells from their average. The statistical uncertainty
of the configuration average \emph{}is again much lower (by a factor
of $\sqrt{35}\approx6$). Hence, statistical noise is not a significant
source of error in these measurements. 

However, the imaging conditions do vary across the field of view,
and can easily differ already in regions separated by a few nanometer
by an amount that is significant for this study. We describe in detail
in the supplementary information how the variation of imaging conditions
across the field of view can be measured for this particular material,
and how differences in imaging conditions between single-layer and
bi-layer reference are minimized. This degree of control for imaging
conditions was not demonstrated in the previous HRTEM studies of this
material \cite{IijimaBN09,ZettlHbn2009}. With a very high precision,
the intensity profiles of the single-layer and double-layer hBN regions
are exactly identical. We estimate an experimental error of 3\% contrast
difference between B and N (relative to the total modulation), based
on comparisons of separated regions with the same thickness. This
result is in agreement with the DFT calculation, but in contrast to
the IAM prediction of 10\% relative contrast difference. In other
words, the ionic character of single layer h-BN is confirmed from
a direct measurement, while the neutral-atom (IAM) charge distribution
can be ruled out.

\section{General remarks and conclusions }

As we have shown, details of the electronic configuration can be detected
in high signal to noise ratio TEM images with carefully controlled
conditions. One key ingredient is the fabrication of samples with
precisely defined geometries \cite{DengMarksBonding06}, and a reduction
of radiation damage in order to allow sufficiently high signal to
noise ratios. Graphene and hBN are special in that they are easy to
prepare in such well defined geometries, and maintain their crystalline
surface configurations even under ambient conditions. We note that
much larger chemical shifts than detected here have been predicted
for a variety of materials in Refs. \cite{DengMarksBonding06,DengGlowingDefects07},
but could not be measured with common TEM sample preparation methods.
The obvious implication is that one would need new ways of sample
preparation and transfer, in order to study these effects in most
other materials.

At the same time, our results also show that bonding effects have
to be taken into account for structural and elemental analysis, in
cases where small differences in elemental contrast are important.
For example, based on the IAM one would assume that the identification
of the B and N sublattices in hBN single layers is possible in bright-field
high-resolution TEM images, already with $2.16\,\textrm{\AA}$ resolution.
However, based on the more accurate DFT potentials, one has to conclude
that this is not possible. The experiment confirms this prediction,
once instrumentation errors are suffiently well adressed. Hence, the
question of which element forms the stable mono-vacancies in h-BN
surfaces under TEM observation \cite{MeyerBNnl09,IijimaBN09,ZettlHbn2009},
remains unclear (a recent scanning-TEM study identified the two sub-lattices,
but did not report the typical defects that are found in TEM observations
\cite{KrivanekBNnature10}). For the case of N doped graphene, the
situation is quite different: Based on IAM, one would expect that
a N substitution in graphene can not be detected. It is exclusively
the effect of binding electrons that makes it possible to see nitrogen
substitutions in graphene. Nitrogen substitutions in graphene - possibly
beam-induced - also provide an alternative explanation to the weak
dark contrast observed in Ref. \cite{MeyerAdatomsNat08}: The DFT
based calculation predicts a ca. 0.6\% dark contrast for the Scherzer
image of the uncorrected microscope, in agreement with the observations
(nitrogen substitutions had been ruled out based on IAM calculations
with atomic configurations from the literature \cite{NsubstDistAndCH05,NsubstCNTbondlength07}).
As a further remarkable point, the contrast of the N defect is primarily
due to a change in the electronic configuration on the neighbouring
carbons, rather than on the nitrogen atom itself. Further, it is due
to a higher moment (a dipole) in the charge on these atoms. Hence,
it is also not possible to model these effects by using modified scattering
factors for partly ionized atoms.

In conclusion, we have shown that it is possible to obtain insights
into the charge distribution in nano-scale samples and non-periodic
defects from TEM measurements with extremely high signal to noise
ratios. For our examples of the nitrogen substitutions in graphene
and hexagonal BN layers, we can assign experimentally observed features
to details in the simulated electron distribution. We can detect a
single light substitution atom in graphene, which is possible only
due to an electronic effect. In the case of hBN, the charge redistribution
leads to a loss of elemental contrast. Instead, the partial ionization
of the material is experimentally confirmed for the single layer.
One key ingredient here is the extraordinary stability of the samples
under the low-voltage electron beam, which allows to obtain extremely
high signal to noise ratios from long exposures. Further, the precisely
defined, ultra-thin sample geometry enables a straightforward analysis.
The DFT based TEM image calculation is irreplacable for the interpretation
of experimental results in these materials, and can provide insights
beyond the structural configurations.

\section*{Author contributions}

J.C.M., A.C. and S.K. carried out TEM experiments. J.C.M., S.K. and
A.C. analyzed the data. S.K. carried out DFT calculations and TEM
simulations based on WIEN2k. A.C. contributed to TEM simulations,
discussions, and analysis. H.-J.P., V.S., S.R. and J.S. developed
the synthesis of N doped graphene. D.K. and A.G. carried out DFT calculations
using VASP. G. A.-S. contributed to TEM simulations. T.I. and U. S.
made Auger spectroscopy measurements. U.K. supervised part of the
work. J.C.M. conceived and designed the study, and wrote the paper.
S.K. and U.K. co-wrote the paper.

\section*{Acknowledgments}

We gratefully acknowledge financial support by the German Research
Foundation (DFG) and the Ministry of Science, Research and the Arts
(MWK) of the state Baden-Wuerttemberg within the SALVE project and
by the DFG within research project SFB 569. T.I. acknowledges the
JSPS Postdoctoral Fellowship for Research Abroad. G.A-S. acknowledges
the support of CONACyT-DAAD scholarship

\bibliographystyle{unsrt}
\bibliography{/home/jannik/publ/bibtex/TEM,/home/jannik/publ/bibtex/2dstuff,/home/jannik/publ/bibtex/books,/home/jannik/publ/bibtex/VASP,/home/jannik/publ/bibtex/BN}

\part*{Supplementary information}

\section{Synthesis of N doped graphene}

Nitrogen doped graphene is prepared by Chemical Vapor Deposition (CVD)
with nickel (Ni) or copper (Cu) as a catalyst. It mostly follows the
procedure which has been developed for pristine graphene and results
in high quality graphene \cite{CVDKimNature09,CVDCopperRuoff09} and
graphene membranes \cite{CVDNiHyeJin10}, with the addition of ammonia
as nitrogen source during part of the synthesis \cite{NdopedGr09}.
The experimental procedure for nitrogen doped graphene growth using
both catalysts is same except reaction time (3 min for Ni and 20 min
for Cu). 

In detail, a Ni or Cu film with 300~nm of thickness was coated onto
a $\textrm{SiO}_{2}/\textrm{Si}$ substrate using an electron beam
evaporator. The metal coated substrate was located on the middle of
a quartz tube and heated to 1000~°C at a 40~°C/min heating rate,
under a flow of argon (500 sccm) and hydrogen (500 sccm). The metal
film on $\textrm{SiO}_{2}$/Si was annealed further at 1000~°C for
20~min. A mixture of gases with a composition ($\textrm{CH}_{4}:\textrm{Ar}:\textrm{H}_{2}$
= 200:500:2500 sccm) was introduced at 980~°C for 3~min (Ni) or
20~min (Cu). Then, the sample was cooled down at a rate of ca. 6.7~°C/min.
During cooling, ammonia was introduced instead of $\textrm{CH}_{4}$
within the temperature range of 800~°C to 400~°C (introducing ammonia
at higher temperatures destroyed the metal layers). The sample is
then further cooled down to room temperature under Ar (200 sccm) and
$\textrm{H}_{2}$ (500 sccm) flow. For transferring the doped graphene
layers onto a TEM grid, poly(bisphenol A carbonate) (PC) was dissolved
in chloroform (solid content : \textasciitilde{} 15 wt\%). The PC/chloroform
solution was spin-coated onto the graphene/metal/$\textrm{SiO}_{2}$/Si
substrate with 500 rpm for 2 min. The PC was deposited onto the substrate
homogeneously with \textasciitilde{}10 \ensuremath{µ}m of thickness.
The graphene/PC film was released free-stranding by chemical etching
of the Ni layer with a $\textrm{FeCl}_{3}$ (1 M) solution, followed
by a concentrated HCl solution. The etching time varied depending
on the substrate size from 30 min to 12 h. The graphene/PC film was
rinsed several times with DI-water and dried with nitrogen gas. It
was then placed onto the TEM grid, and the PC was dissolved using
chloroform. Before inserting into the microscope, in order to further
reduce adsorbed contamination, the TEM grid with graphene layer was
heated in air to 200 °C for 10 minutes.

\begin{figure}
\includegraphics[width=0.5\linewidth]{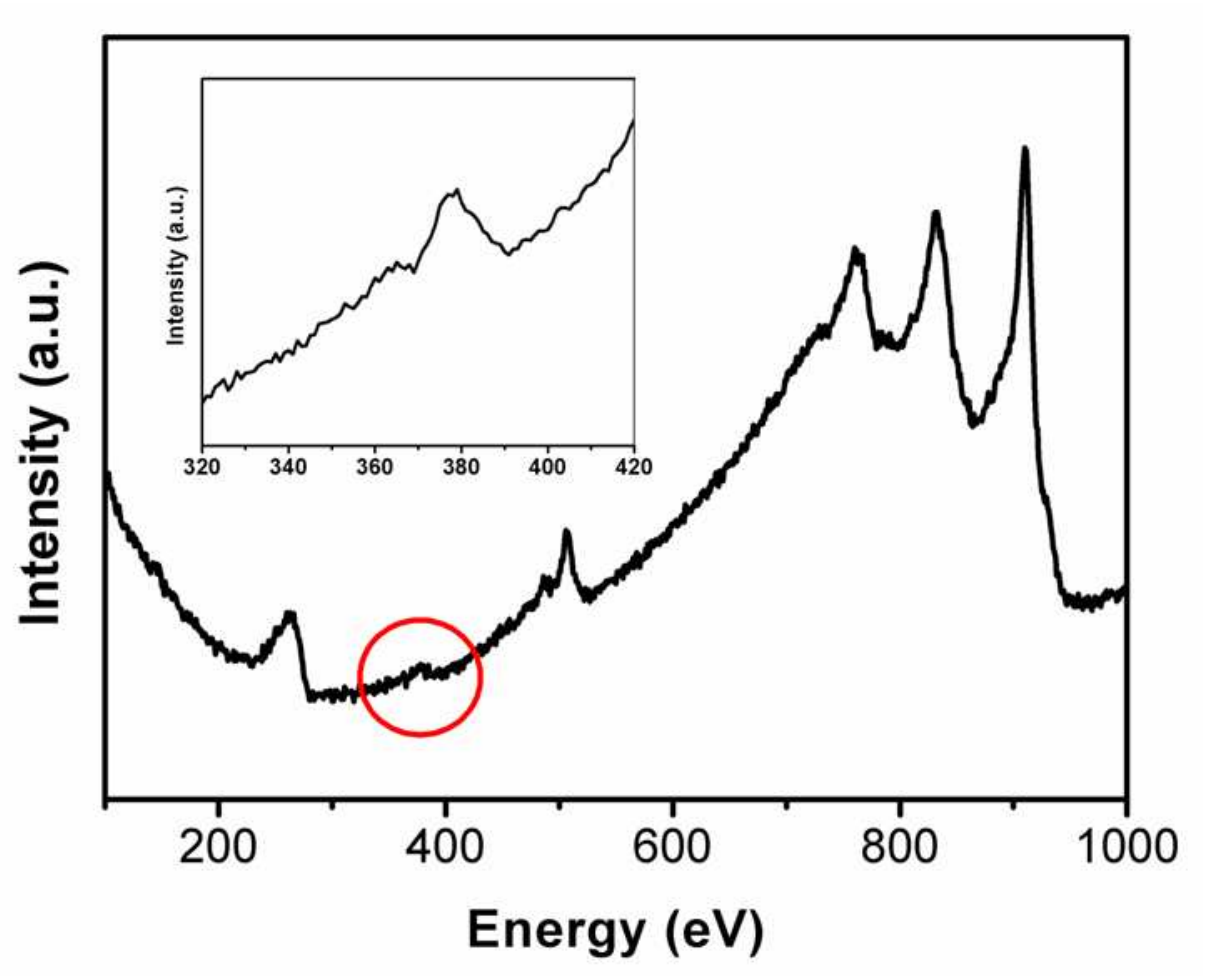}

\caption{Auger spectrum showing the existence of nitrogen in a graphene sheet,
grown on a Cu substrate by the CVD method with the addition of ammonia.\label{fig:Auger}}
\end{figure}

The presence of nitrogen in the doped samples was verified by Auger
spectroscopy (Fig. \ref{fig:Auger}). A small signal at the N KLL
energy is found (enlarged as inset in Fig. \ref{fig:Auger}), which
is not present in undoped reference samples. The other lines in the
spectra correspond to carbon, oxygen (presumably physisorbed after
air exposure) and copper, and are also found in the undoped graphene
reference samples. Auger spectroscopy was performed using a JEOL JAMP-7810
scanning Auger microprobe at 15keV. 

\section{Additional images for N doped graphene}

\begin{figure}
\includegraphics[width=0.84\linewidth]{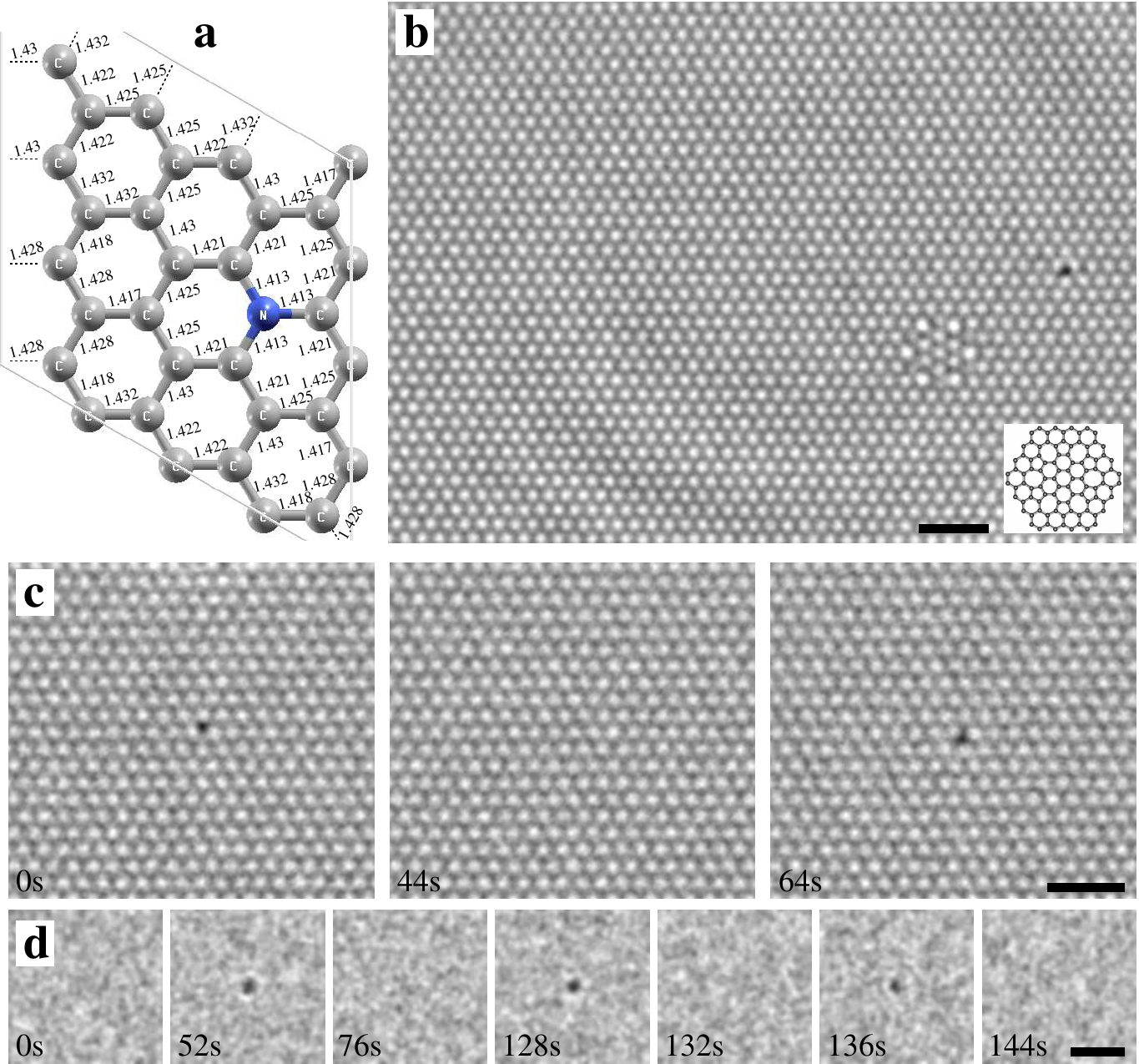}

\caption{Nitrogen doped graphene. (a) DFT relaxed configuration of a single
nitrogen substitution in graphene, with bond lengths given in $\textrm{\AA}$.
 (b) Single-layer graphene membrane with the flower defect. Inset
shows a structural model. (c,d) Sequences showing repeated trapping
and detachment of contamination atoms on the nitrogen substitutions.
 Sequence (c) is recorded at Scherzer defocus (ca. 10nm) and (d) at
an underfocus of ca. 18nm, with the lattice contrast removed by a
Fourier filter. All scale bars are 1~nm. \label{fig:Sngr}}
\end{figure}

Fig. \ref{fig:Sngr}a shows the DFT relaxed configuration of N doped
graphene as used in all simulations. Fig. \ref{fig:Sngr}b-d shows
some additional observations from nitrogen doped graphene. Fig. \ref{fig:Sngr}b
shows an example of an unexpected defect configuration that was frequently
found in the nitrogen doped samples that were synthesized on the Ni
surface. It consists of seven hexagons that are rotated by 30°, connected
to the graphene lattice by six pentagons and heptagons. Due to its
peculiar appearance we will refer to it as flower defect. Its formation
mechanism remains unclear at this time. Remarkably, the flower defect
is not a reconstructed vacancy; i.e., it contains the same number
of atoms as the ideal graphene sheet. 

We note that the nitrogen defect is reactive \cite{NsubstDistAndCH05},
at least in comparison to the inert carbon surface. Frequently, the
nitrogen substitution defects have isolated ad-atoms trapped on top
at the beginning of the observation, which then detach under electron
irradiation after a few exposures and thus reveal the isolated nitrogen
substitution defects. These ad-atoms are trapped on the nitrogen atom
or its neighbouring carbons. In some cases, we observe a repeated
trapping and de-trapping of contamination atoms on the nitrogen defect
in long image sequences. The sequence in Fig. \ref{fig:Sngr}c is
recorded at Scherzer defocus and shows direct structure images. An
ad-atom is present initially, desorbs in intermediate images, and
another ad-atom is again trapped in the same place (within one nearest
neighbour). Sequence (d) is recorded at the larger defocus (-18nm,
optimized for N substitution detection), the lattice is filtered out.
Here, several events are observed where an ad-atom is trapped on the
same position. Indeed, this dynamics provides an independent confirmation
of the nitrogen defect in this position: The repeated trapping of
ad-atoms in the same position (within one nearest neighbour) proves
that there must be some kind of reactive defect within the single
layer graphene sheet at this point. However, the intermediate frames
appear identical to undisturbed graphene within the noise level of
single exposures, which allows only substitutions of similar atomic
number (B, N, O) as alternatives to carbon. Then, given the intentional
presence of nitrogen during synthesis of these samples, we can identify
these defects as nitrogen substitutions.  For the nitrogen substitution
analysis as shown in the main article, we have to use averages from
several exposures. For this purpose, we only use parts of the image
sequences where no ad-atom is visible in all individual exposures.

\begin{figure}
\includegraphics[width=1\linewidth]{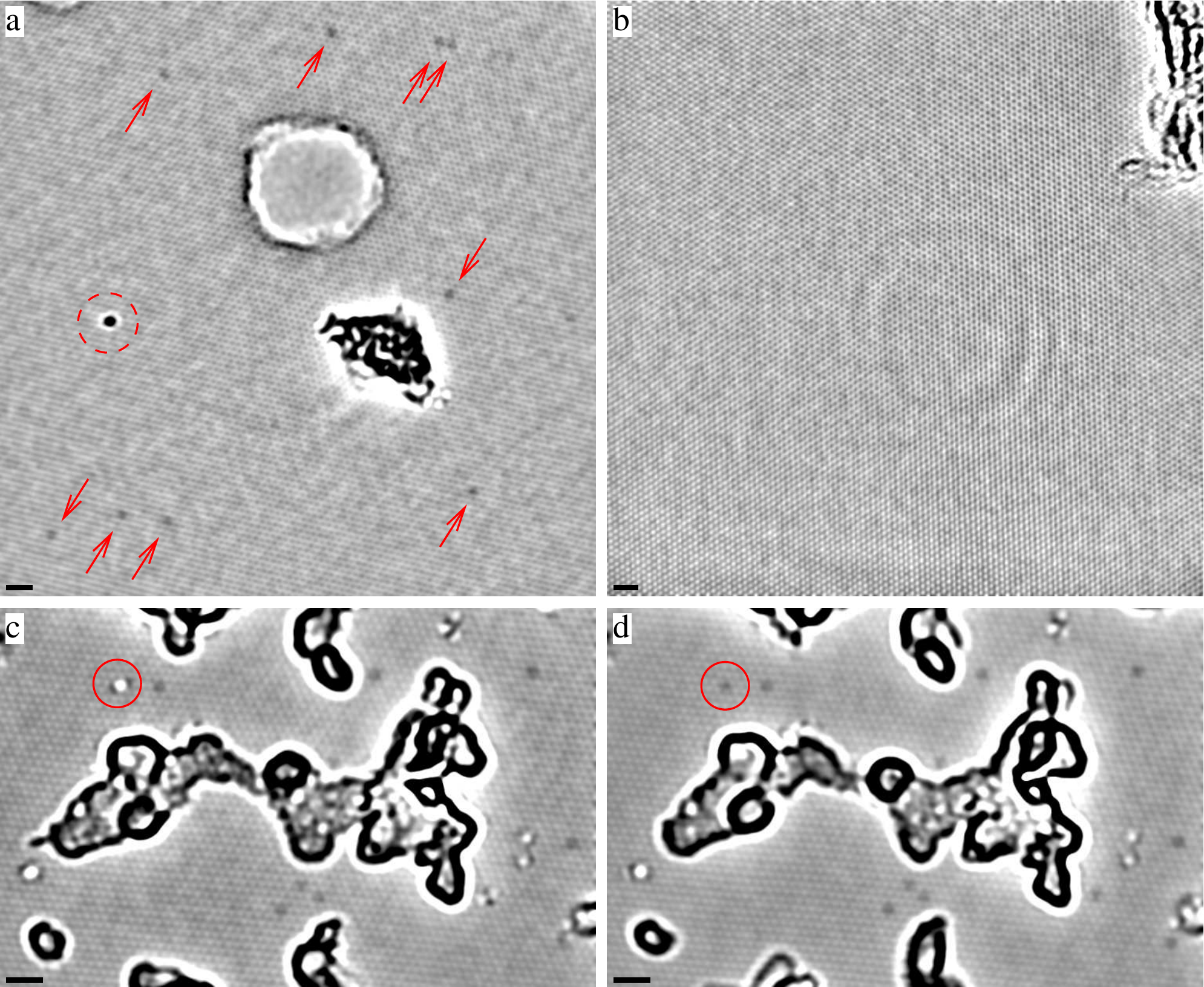}

\caption{Further images of nitrogen substitutions in graphene. (a,b) Comparison
of a nitrogen doped sample (a), and an un-doped graphene sample (b).
9 nitrogen substitutions are visible as weak dark spots and indicated
by red arrows in (a). The heavy dark spot (red dashed circle) is most
likely an ad-atom trapped on a nitrogen defect (see Fig. \ref{fig:Sngr}c,d).
(c,d) Irradiation-induced nitrogen substitutions in two subsequent
images. In one case, the transition from a vacancy-type defect to
a nitrogen substitution was observed. All scale bars are 1nm.\label{fig:FurtherNimg}}
\end{figure}

In Fig. \ref{fig:FurtherNimg} we show further TEM images of nitrogen
substitutions in graphene. Fig. \ref{fig:FurtherNimg}a,b shows the
comparison of a nitrogen doped sample and an un-doped reference graphene
sample. In this case, Fig. \ref{fig:FurtherNimg}a shows a graphene
sheet grown on a Ni surface in presence of ammonia, and Fig. \ref{fig:FurtherNimg}b
shows a graphene sample that was prepared by mechanical cleavage \cite{MeyerEBIDapl08}
(and not irradiated above the knock-on threshold \cite{KnockOnDamageJAP01}).
Fig. \ref{fig:FurtherNimg}c,d shows a sample that was also prepared
by mechanical cleavage, briefly irradiated with 300keV electrons (with
a dose of ca. $10^{7}\frac{e^{-}}{\textrm{nm}^{2}}$). The sample
shown in Fig. 2i of the main article was left in the microscope vacuum
between irradiation and imaging (ca. 2 hours later), while the sample
in Fig. \ref{fig:FurtherNimg}c,d was exposed to air. In both cases,
we observe nitrogen substitution defects. A likely mechanism is a
substitution of irradiation-induced vacancies by nitrogen atoms. Indeed,
the sequence in Fig. \ref{fig:FurtherNimg}c,d shows one example where
a vacancy defect converts into a nitrogen substitution defect during
observation. All images in Fig. \ref{fig:FurtherNimg} are averages
of ca. 40 exposures, recorded at a defocus of $f_{2}\approx-18\textrm{nm}$
and a gaussian low-pass filter was applied such that the graphene
lattice is almost suppressed but still discernible. Also, we note
in the larger-area views (Fig. \ref{fig:FurtherNimg}a,b) that the
imaging conditions are close to ideal only near the amorphous adsorbates
that are used for fine tuning. The contrast of the nitrogen substitution
is only weakly dependent on small residual non-round aberrations.
Therefore, this topic is discussed in more detail below in the context
of hexagonal boron nitride images, where it is important for the analysis.

\section{Details of the VASP DFT simulation}

Spin-polarized DFT calculations were carried out with the Vienna ab
initio simulation package (VASP)~\cite{Kresse96}. The structure
of the substitution defects in graphene was relaxed using the exchange-correlation
functional by Perdew, Burke and Ernzerhof (PBE)~\cite{Perdew96}.
Ionic cores were represented by the projector augmented-wave (PAW)
method~\cite{Bloechl94PAW,Kresse99PAW}. An energy-cutoff of 400
eV was used for the calculations. With these parameters, the pure
graphene layer has a relaxed C-C-distance of 1.424\,{\AA}. For
the calculations of the defect structure, a $4\times4$ supercell
of the hexagonal graphene unit cell was used (lattice constant: 9.868\,{\AA}).
We employed a large $9\times9\times1$ k-point mesh with a Gaussian
smearing of 0.1\,eV for the k-point summation in order to ensure
well-converged geometries of the considered structures. Different
initial configurations with the nitrogen atom slightly displaced in
different directions all resulted in the same relaxed structure. The
relaxation process was stopped when all forces were smaller than 0.002
eV {\AA}$^{-1}$.

\section{Details of WIEN2K DFT simulation}

The all-electron WIEN2K DFT simulation is used to obtain the electron
charge density and electrostatic potential, for a structure with fixed
atomic coordinates. For boron nitride we use the literature value
of $a=2.505\,\textrm{\AA}$ \cite{BNlatticeconst2505Solozhenko1997}
(separation between B and N atoms of $a_{BN}=1.45\,\textrm{\AA}$),
and for nitrogen doped graphene we use the relaxed configuration described
above (see Fig. \ref{fig:Sngr}a). We have evaluated the required
number of k-points and size of the base set (RKMAX) by convergence
tests. To this end, we have not only tested a convergence of the total
energy and electric field gradient, but also of our value of interest,
the projected potential values. For the results shown in the main
article, we used 54 k-points with 15x15x3 division for the hBN results.
For the nitrogen substituted graphene supercell, we used 8 k-points
with 4x4x1 division and RKMAX=8\emph{.} The muffin-tin radius is set
automatically by WIEN2K to create spheres that touch with the nearest
neighbours. All other values are left to the defaults of the WIEN2K
software (version 9.1).

\section{Image simulation based on WIEN2K results}

\subsection{Overview}

For the TEM image simulation, it is important that all electrons are
included in the simulation, and not only valence electrons. For this
reason, we extract the electrostatic potentials from the WIEN2K results.
This software was also used by Deng and Marks in Ref. \cite{DengMarksBonding06}.
However, they used X-ray scattering factors calculated by WIEN2K and
transformed them to electron scattering factors using a modified version
of the Mott-Bethe formula \cite{DengPengCowleyMottForm88}, while
we extract potentials directly from WIEN2K. The advantage of our approach
is the simplicity and in principle the possibility to carry out multi-slice
simulations (not needed for our samples), while the drawback is a
very fine required sampling of the potential and associated high computing
time. The WIEN2K software calculates the electrostatic potential created
by all electrons and all core charges. This potential is stored in
a file called case.vcoul and contains the Coulomb potential in a lattice
harmonics representation and as Fourier coefficients. The utility
program lapw5 can be used to obtain linescans and 2d slices of this
potential file. A large number of 2d slices is then extracted to obtain
the 3d potential. We then calculate the projected potential in the
direction of view of the TEM (i.e., normal to the graphene or hBN
layers), and simulate the effect of the microscope as described further
below.

\subsection{Electrostatic potentials}

\begin{figure}
\includegraphics[width=1\linewidth]{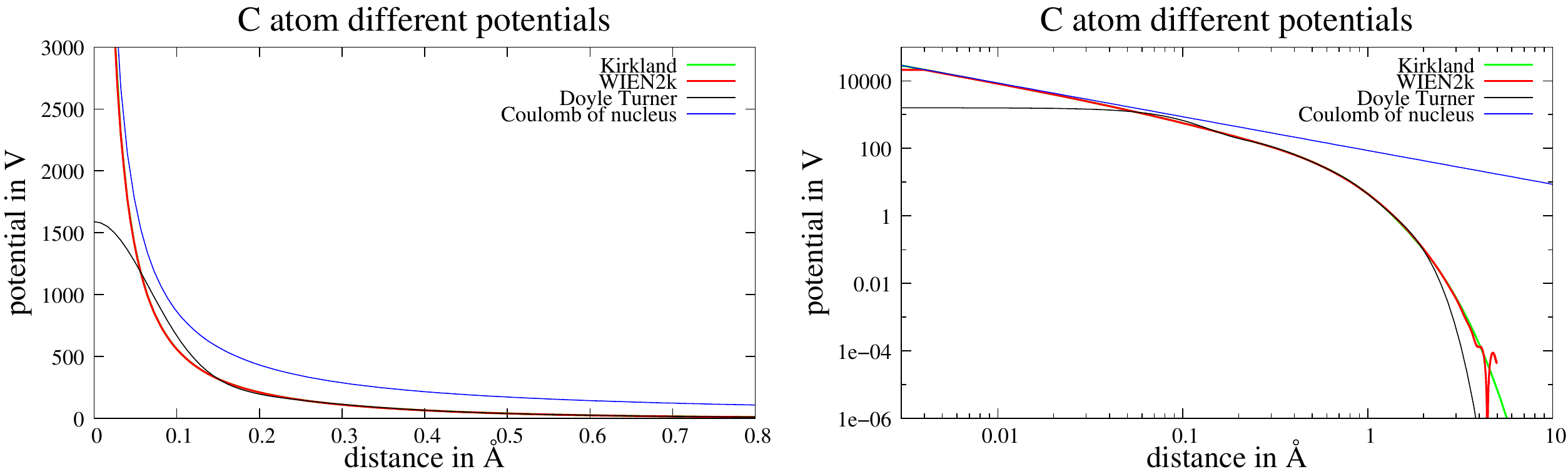}

\caption{Carbon atom electrostatic potentials, using the parametrization of
Doyle and Turner \cite{DoyleTurner68}, the parametrization of Kirkland
\cite{KirklandEMcomputing}, the electrostatic potential extracted
from WIEN2K for an isolated atom, and the unscreened Coulomb potential
of the nucleus for comparison in a linear and double logarithmic plot.\label{fig:Cpot}}
\end{figure}

The aim of the next section is to clarify that the potential extracted
from WIEN2K can be used directly for TEM image simulation. As a first
test, we compare the results for an isolated atom with existing isolated-atom
potentials. In Fig. \ref{fig:Cpot} we see the comparison between
the WIEN2K starting potential and two different isolated atom potentials
by Doyle and Turner \cite{DoyleTurner68} and Kirkland \cite{KirklandEMcomputing}.
The WIEN2K potential was obtained by defining a rectangular unit cell
of $10\times10\times10\,\textrm{\AA}$ with a single carbon atom inside.
The starting charge density and the corresponding potential files
were calculated and a linescan with a resolution of $250\,\textrm{pp\AA}$
was created using the utility program lapw5. Far from the core the
WIEN2K potential becomes constant but not zero. The linescans were
shifted in z-direction (until the smallest value was equal zero) to
obtain the WIEN2K potential in the common normalization. One factor
that complicates this normalization is the fact that the WIEN2K potential
starts to oscillate in the vacuum region, which is probably caused
by the limited size of basis functions. However, the oscillations
and deviations occur only in vacuum regions where the potential is
very small and therefore do not affect our results. From the graphs
in Fig. \ref{fig:Cpot}, we see that the WIEN2K and Kirkland potential
agree very well in the core region where, in contrast to the Doyle
Turner potential, both of them are divergent.

The excellent agreement of the isolated-atom potential from WIEN2K
to the Kirkland potentials shows that the WIEN2K potentials are well
suited for TEM simulation, and small changes from bonding should not
affect this point. As a further useful fact, we found that the starting
potentials from WIEN2K (i.e., results obtained with zero iterations
of the DFT computation) are practically indistinguishable from the
Kirkland potentials. In other words, the WIEN2K DFT simulation uses
the independent-atom electron distribution as starting configuration.
This provides an easy way to obtain an independent-atom (conventional)
TEM simulation result for comparison with the DFT based result: Simulated
images based on potentials extracted with zero iterations in WIEN2K
correspond to the conventional (IAM) TEM simulation, while bonding
effects are taken into account after the DFT based changes in the
electronic configuration have been computed. In this way we can compare
IAM and DFT results in a way that uses precisely the same algorithms
and settings for modeling the effect of the microscope.

An important point that remains in working directly with the electrostatic
potential in an equidistant 3-d grid is that we are making a discretization
of a function that has poles at the positions of the nuclei. In this
case, sampling points around the nucleus may not adequately represent
the integrated potential of the corresponding volume element. In the
extreme case, a sampling point may fall onto the position of the nucleus
and thereby contribute an infinite value, multiplied with a finite
volume element. To overcome this problem, we sample the electrostatic
potential at a much higher spatial resolution than needed for image
simulation, and use a global cut-off parameter for the potential to
avoid infinite contributions. In other words, we mitigate the effect
of the {}``bad pixel'' at the position of the nucleus by using a
very fine sampling, and subsequent averaging (this averaging or smoothing
is inevitably part of the TEM simulation for the much lower realistic
experimental resolution in the following step). Then, the main question
is what sampling (points per $\textrm{\AA}$) is needed, and how much
error one has to expect as a result. To get an idea of the numbers,
sampling at 30 points per $\textrm{\AA}$ and required (experimental)
resolution of only $\approx1\,\textrm{\AA}$ means that there are
$30^{3}=27000$ volume elements within a $1\,\textrm{\AA}$ cube,
of which only one, which includes the nucleus, is affected by the
cutoff. In this respect, it is important to note that the cut-off
is implemented on the 3-D potential, and not the (2-D) projected potential
which serves as input for the simulation. 

\begin{figure}
\includegraphics[width=1\linewidth]{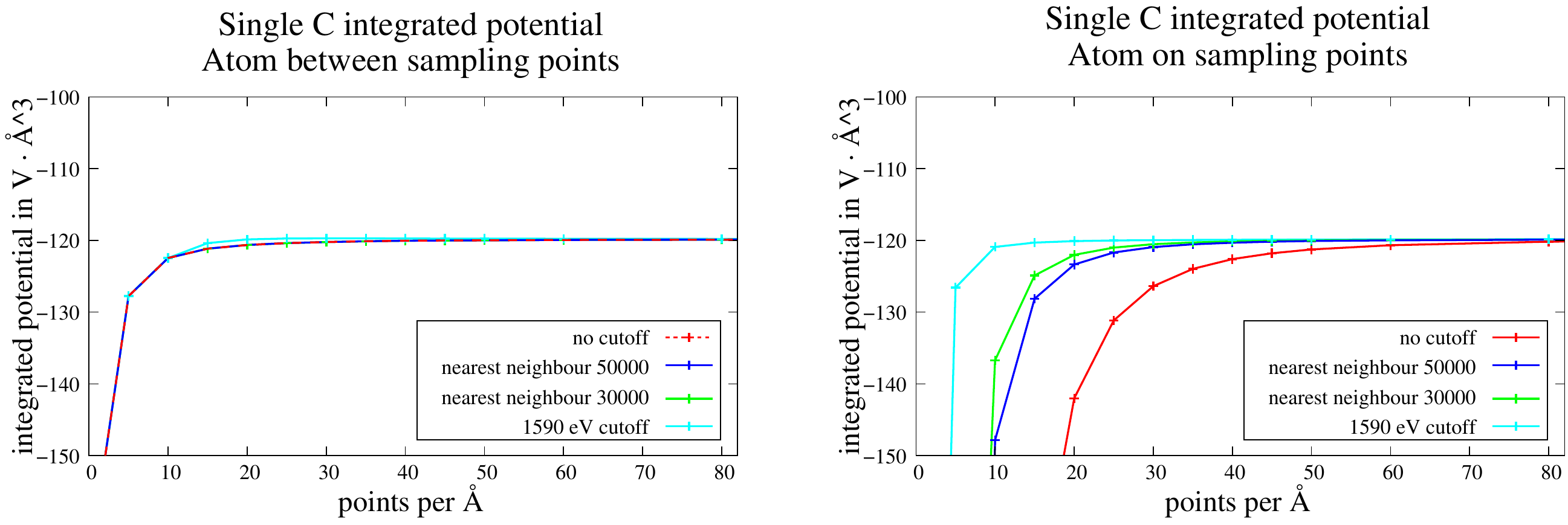}

\caption{Integrated potential of a single carbon atom, extracted from the
WIEN2K electrostatic potential using different numerical sampling
and different cutoff values for the singularity at the nucleus. Also,
the atom was placed halfway in-between sampling points (left) and
onto a sampling point (right). At a sufficiently high sampling rate,
all results converge to the same value. \label{fig:convtest}}
\end{figure}

The required number of points is determined in a convergence test,
where we use the integrated potential of the atom as the figure of
merit. This is reasonable since, at the rather low experimental resolution
(compared to the sampling in the simulation), only the integral value
from the core region of the atom is relevant. In Fig. \ref{fig:convtest},
we show the integrated potential for an isolated atom after extracting
the potential with different sampling and different cutoff values.
Also, the nucleus was placed onto the sampling point and in-between
for comparison. As cut-off method, we use a nearest-neighbor cutoff,
meaning that the potential value is reduced to that of the nearest
pixel if it exceeds 30000 or 50000 volts, and a fixed cutoff of 1590V
(maximum of the Doyle-Turner potential for carbon). If we use no cut-off,
one would expect an infinite value if the atom is on the sampling
point (an infinite value multiplied with a finite volume element numerically)
but the maximum output value from lapw5 is $1.6\cdot10^{6}\, V$ (i.e.,
effectively another cutoff due to the numerical value range). Even
with this extreme variation of values for the volume element that
includes the nucleus, the integrated potential precisely arrives at
$(119.8\pm0.1)\, V\cdot\textrm{\AA}^{3}$ for the highest sampling.
In other words, the value of this one volume element does not matter
if simply a sufficiently high sampling is used. The value is also
in a very good agreement with the integrated potential from the Doyle-Turner
parameters \cite{DoyleTurner68} of $120.03\, V\cdot\textrm{\AA}^{3}$
(which can be integrated analytically). Further, we can conclude that
a sampling of 25-35~pp$\textrm{\AA}$ is sufficient for our purposes
if we use a reasonable cut-off algorithm. For the results presented
here, we use a sampling rate of 104~pp$\textrm{\AA}$ (hBN) and 52~pp$\textrm{\AA}$
(N doped graphene unit cell) with the 50~kV nearest-neighbour cutoff. 

As a further cross-check, we note again that we can obtain both the
conventional (IAM) and DFT-based TEM simulation from the starting
and final electron distribution in WIEN2K. For both cases, we use
precisely the same sampling, integration, and positions of the nuclei
with respect to the sampling points. Effects from poor sampling of
the potential close to the nucleus, if present, would occur in both
cases with the same magnitude (the positions of the nuclei are the
same, and the electron charge density does not contain singularities).
Hence, any sampling effects would show up in the comparison of our
IAM result to other IAM simulation codes, based e.g. on Doyle-Turner
potentials (which avoid the singularity), and this is not the case.

\subsection{TEM image simulation}

The previous section described how to obtain accurate electrostatic
potentials on the atomic scale for a given structure in a way that
includes the effects of chemical bonding. The first part of a TEM
image simulation is now to model the effect of the sample on the electron
beam. Since our samples consists of a single atomic layer, we use
a single projection of the electrostatic potential. Then, the effect
of the sample is simply a phase shift in the electron wave which is
proportional to the projected electrostatic potential

\[
\Psi(x,y)=\Psi_{0}(x,y)*\exp(-i\sigma V_{z}(x,y))\]

where $\Psi_{0}(x,y)$ and$\Psi(x,y)$ is the wavefunction before
and behind the sample, respectively,\[
V_{z}(x,y)=\int V(x,y,z)dz\]

is the projected potential, and $\sigma=\frac{2\pi me\lambda}{h^{2}}$
the interaction parameter (where both electron mass $m$ and wavelength
$\lambda$ depend on the electron energy; $e$ is the electron charge,
$h$ is the Planck constant). The projected potential is calculated
numerically from the 3-D potential obtained from WIEN2K as described
above.

\begin{figure}
\includegraphics[width=0.4\linewidth]{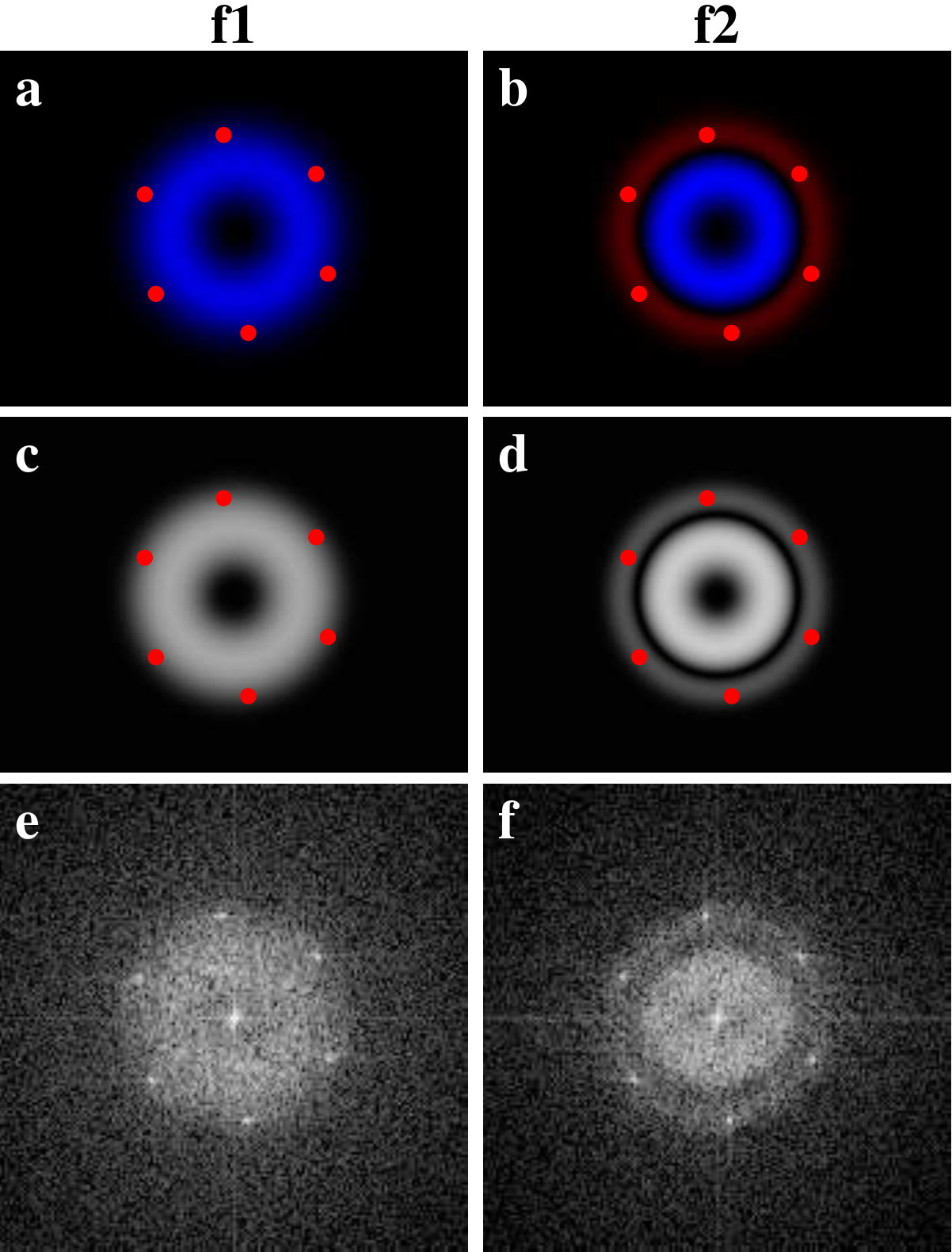}

\caption{Contrast transfer at the conditions used in this work. (a,b) Phase
plate $\sin(\chi(q_{x},q_{y}))E(q_{x},q_{y})$, for focus $f_{1}$
and $f_{2}$ as used for the simulations. Blue and red represents
positive and negative values, respectively. (c,d) Square modulus (logarithmic
greyscale) of the phaseplates. Red dots in (a-d) indicate the $2.13\textrm{\AA}$
$(1-100)$ graphene reflections. (e,f) Power spectra (log scale) of
the experimental images from a small amorphous adsorbate on the graphene
sheet close to our region of interest. \label{fig:CTF} }
\end{figure}

The second part of the TEM image simulation is to model the effect
of the microscope, which follows the standard approach as in Chapter
3 of Ref. \cite{KirklandEMcomputing}. We use a spherical aberration
of 20\ensuremath{µ}m, an acceleration voltage of 80kV, a focal spread
of 5nm, and a convergence angle of 0.5~mrad. The two defocus values
are $f_{1}=-90\,\textrm{\AA}$ (Scherzer defocus) and $f_{2}=-180\,\textrm{\AA}$
(graphene lattice in the second extremum of the contrast transfer
function (CTF)). The corresponding phase plates (including damping
envelopes $E(q_{x},q_{y})$) are shown in Fig. \ref{fig:CTF}a,b.
Experimentally, imaging conditions can be verified from the thin amorphous
adsorbates that cover some parts of the graphene sheet. Fig. \ref{fig:CTF}e,f
shows the Fourier transform from such adsorbates, recorded in the
same images and only a few nm away from our features of interest.
For focus value $f_{1}$, the single, continuous pass-band with the
graphene lattice spacing at the outer rim is clearly visible in Fig.
\ref{fig:CTF}a,c,e. For focus value $f_{2}$, a single dark ring
in the power spectrum, just inside the graphene lattice spacing, is
expected and confirmed in the experimental data (Fig. \ref{fig:CTF}b,d,f).

\section{TEM Data acquisition and processing}

TEM imaging was carried out using an image-side spherical-aberration
corrected Titan 80-300, operated at 80~kV, with the exception of
Fig. \ref{fig:FurtherNimg}c,d (50~kV) and Fig. \ref{fig:bn100}
(100~kV). The extraction voltage of the source was reduced from its
standard value (ca. 4~kV) to 2~kV, in order to obtain a lower energy
spread. Images were recorded on the CCD camera (1k x 1k, Gatan MSC
742) or the CCD camera within the Gatan Imaging Filter (GIF), a 2k
x 2k camera (Model US1000). The GIF camera was operated in a 2x binning
mode in order to obtain a faster read-out; i.e. we were effectively
using it as another 1k x 1k camera. In both cases, we use a sampling
of 5 pixels per Angstrom (referring to the binned pixel size for the
GIF camera).

Image sequences are recorded with ca. $10^{4}$ counts per pixel,
1 second exposure time, and ca. 2 second intervals. Effects of slightly
uneven illumination are removed by normalization (division) to a copy
of each image with a very large Gauss blur ($>40\textrm{\AA}$ FWHM)
applied. In this way, the long-range variations of the uneven illumination
are removed, and at the same time the mean intensity is normalized
to 1 throughout the image. The image data is upscaled to a 2x higher
sampling ($0.1\textrm{\AA}$ pixel size) prior to image alignment,
to avoid interpolation artefacts. Image registration is done with
the Stackreg-plugin \cite{IJstackreg} for the ImageJ software, which
provides sub-pixel accuracy for image alignment. In addition, Fig.
5a of the main article was rotated to be aligned with the crystal
orientation.

\section{Additional calculation results for nitrogen doped graphene}

\begin{figure}
\includegraphics[width=1\linewidth]{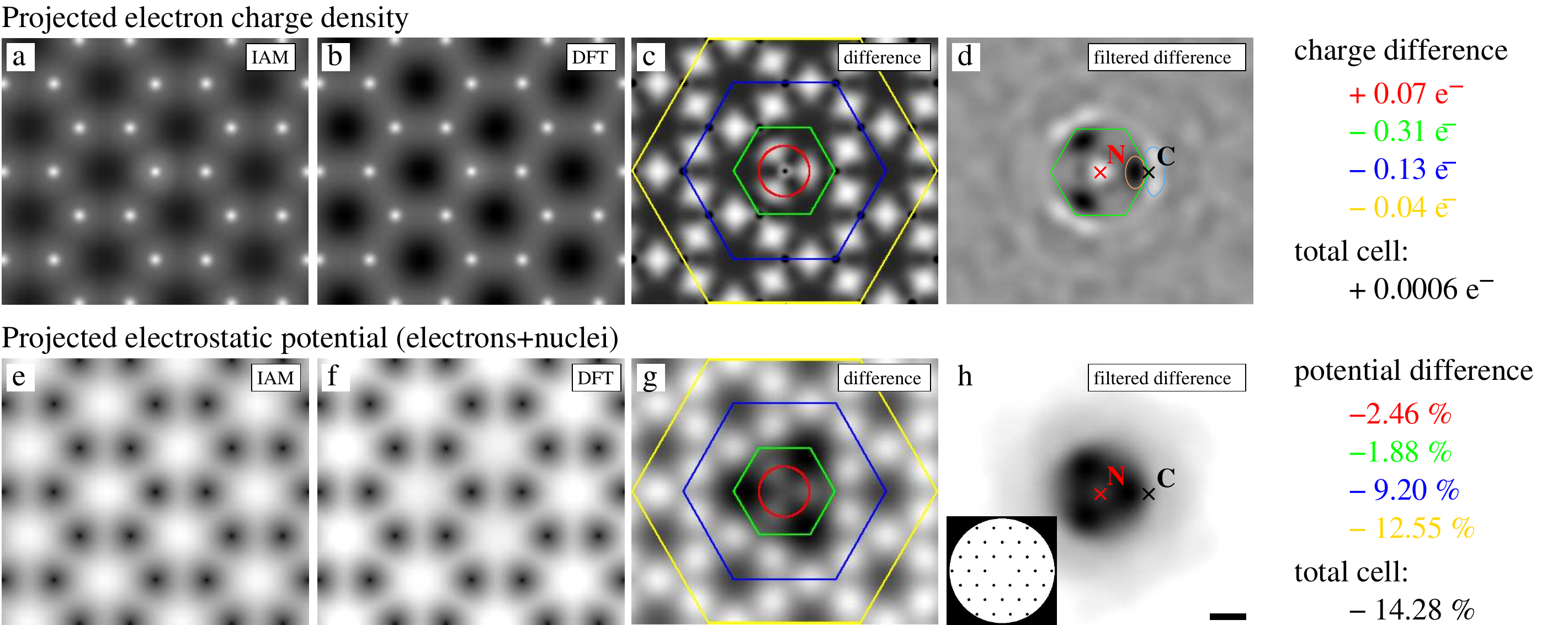}

\caption{Analysis of the nitrogen defect calculation results. (a-d) Analysis
of the projected electron density, (a) IAM and (b) DFT result (log
greyscale, white corresponds to a higher charge density). (c) Difference
DFT-IAM, showing charge rearrangement due to bonding (linear greyscale,
white corresponds to a higher charge density in the bonded configuration).
Charge differences in the marked regions is given on the right hand
side. (d) Charge density difference (DFT-IAM), with periodic components
of the graphene removed. The dipole-shaped charge on the carbon atoms
next to the nitrogen atoms is indicated in one example as orange-blue
line. (e-h) Analysis of the projected potentials, (e) IAM and (f)
DFT result (log greyscale). (g) Difference DFT-IAM (linear greyscale).
Dark corresponds to high projected potential values. Relative differences
listed on the right hand side are (DFT-IAM)/DFT. (h) Difference in
projected potentials, with periodic components removed (mask used
in d,h is shown as inset). Scale bar is $1\textrm{\AA}$.\label{fig:AndSim}}
\end{figure}

Although the TEM image depends directly on the electrostatic potentials,
it is instructive to look at both, the projected potentials and the
charge densities for IAM and DFT calculation. These two properties
are related to each other via Poisson's equation in the three-dimensional
volume (however, note that we are showing the projected charge densities
and potentials here). The difference between the IAM and DFT results
shows how charges are re-arranged upon bond formation. Fig. \ref{fig:AndSim}a
and b show a projection of the charge density, for the IAM and the
DFT calculation, respectively. The change from a spherical charge
distribution (IAM, Fig. \ref{fig:AndSim}a) to the charge distribution
that is smeared out over the bonds (Fig. \ref{fig:AndSim}b) is apparent
in the comparison of these two images. Fig. \ref{fig:AndSim}c shows
the difference in the projected electron charge densities. In order
to estimate a degree of ionization, it is necessary to define a volume
surrounding each atom (i.e., an area in the projected charge density),
and any choice here is of course arbitrary. First, we measure the
excess charge on the nitrogen atom itself, by choosing a surface that
cuts between the C and N atom (red circle in Fig. \ref{fig:AndSim}c).
Here, the change is only $+0.07e^{-}$as compared to a neutral atom
(in agreement with a previous calculation \cite{NsubstDistAndCH05}).
Remarkably, the nitrogen atom itself is almost neutral and in fact
is charged with the wrong sign to explain a dark contrast. Also, all
carbon atoms are nearly neutral when choosing such a cylindrical boundary
around the atoms. However, a hexagon shaped boundary through the nearest
carbons (yellow in Fig. \ref{fig:AndSim}c) shows a difference of
$-0.31e^{-}$ in this larger region (including the central circle).
This reduced electron density, or positive net charge (reduced screening
of the core potential) spread over a larger area, explains the smooth
dark contrast associated to the nitrogen substitution. This charge
is then mostly screened over a further atomic distance (blue hexagon).
Further insight can be gained by removing the periodic components
of this difference image, shown in Fig. \ref{fig:AndSim}d: From this
image, it is clear that the most dominant effect is a dipole-shaped
rearrangement of the electrons on the nearest-neighbour carbons around
the nitrogen defect. Again, this delocalized charging effect agrees
with the smooth appearance of the nitrogen defect in graphene as predominantly
lower frequency components in the high-resolution experimental images.
We can thus conclude that, in our experiment, the nitrogen substitution
can be detected and appears the way it does mostly because of a charging
effect on its neighbouring carbon atoms.

A similar analysis is shown for the projected potentials in Fig.\ref{fig:AndSim}e-h.
Here, the potential differences are given as relative differences
(DFT-IAM)/DFT. In the comparison of the projected potentials, it should
be noted that already for ideal graphene, the mean inner potential
is reduced by 15\% in the DFT result as compared to the IAM. The important
insight here is that we obtain an \emph{extended} region around the
nitrogen defect where the projected potential is higher than in the
surrounding graphene, visible in Fig. \ref{fig:AndSim}g,h as extended
dark area.

\section{Additional calculation results for hexagonal Boron Nitride}

\begin{figure}
\includegraphics[width=0.64\linewidth]{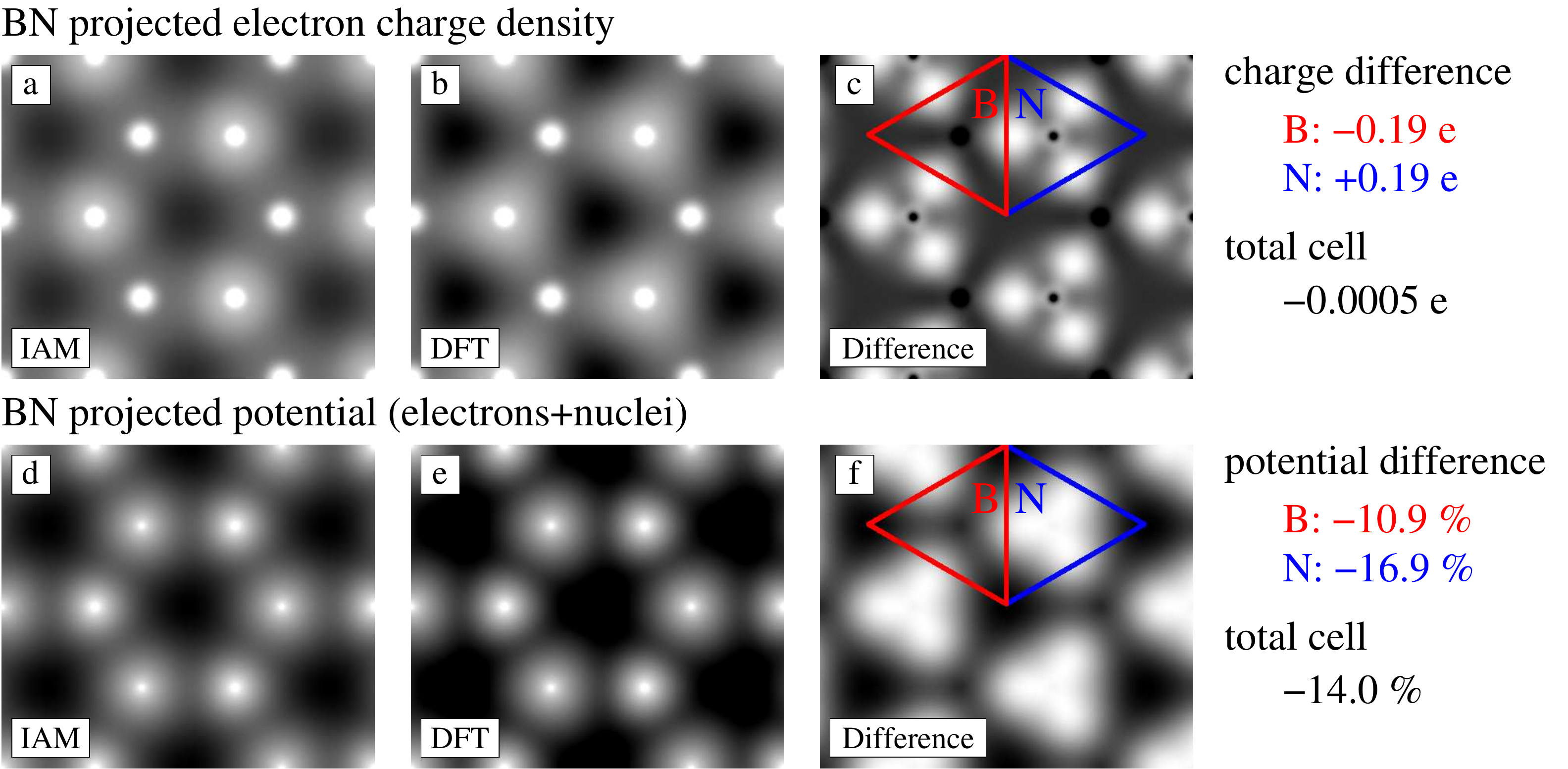}

\caption{Projected electron charge densities and projected potentials for
single layer h-BN for the IAM and our DFT simulation. White corresponds
to a high electron density or high projected potential. (a,b,d,e)
are represented on a logarithmic greyscale, while (c,f) use a linear
greyscale. (a,b) projected electron charge density for IAM and DFT,
respectively. The change from a roundish shape of the electron cloud
(a) to a triangular shape (b) due to bonding effects is visible. (c)
Differences (DFT - IAM) in the projected charge density, i.e., charge
shifts due to bonding. Using the half-unit-cell boundary as indicated,
a shift of 0.2 electrons from B to N is found. (d, e) projected potentials
for IAM and DFT. In the DFT result (e), the core potentials are screened
over a shorter distance, resulting in a {}``sharper'' appearance
of the atoms as compared to (d). Correspondingly, the mean inner potential
of the whole unit cell is decreased by 14\% in the DFT result. (f)
Differences in the projected potential, integrated over the indicated
areas. Relative differences are given as (DFT-IAM)/(IAM).\label{fig:BNanalysis}}
\end{figure}

We analyze again both the projected potential and charge distribution,
shown for hBN in Fig. \ref{fig:BNanalysis}. In the charge density,
the change from a roundish (IAM, Fig. \ref{fig:BNanalysis}a) to a
triangle-shaped distribution (DFT, Fig. \ref{fig:BNanalysis}b) is
apparent in the comparison. Fig. \ref{fig:BNanalysis}c shows the
difference in the projected electron charge densities. If we split
the unit cell into two equal areas as indicated by the triangles in
Fig. \ref{fig:BNanalysis}c, we find that 0.2 electrons shift from
the boron to the nitrogen atom due to bonding (note that this value
of partial ionization depends strongly on how these volumes are defined).
In the projected potentials, we find that already the mean inner potential
(the volume average of the potential) is 14\% lower in the DFT result
as compared to IAM (this difference might be detectable e.g. by electron
holography). In the DFT result, the core potentials are screened over
a shorter distance than in IAM, resulting in a slight reduction of
atomic contrast and a sharper appearance of the atoms. However, this
effect is stronger on the N site, leading to a practically cancelled
contrast difference between B and N at our experimental conditions.
The accumulation of charge on the nitrogen site is in agreement with
the analysis of h-BN in the literature \cite{PaulingBNandGrPNAS1966}.
Again, the experimental TEM data confirms this ionic character of
h-BN for the single layer, and rules out the neutral atomic configuration.

\begin{figure}
\includegraphics[width=0.72\linewidth]{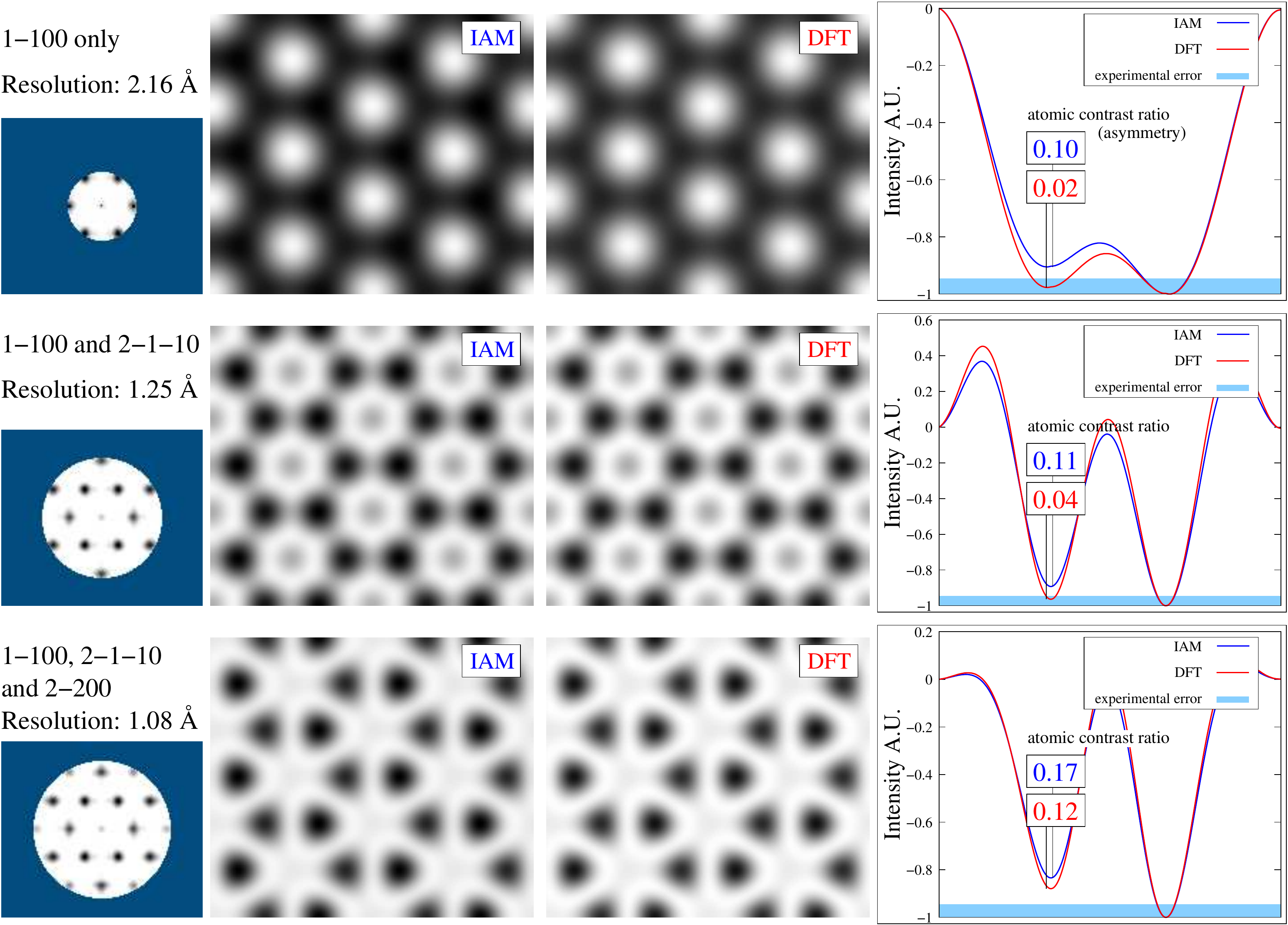}

\caption{Simulations from IAM and DFT for different resolution. Shown is the
projected potential, bandwidth-limited by different apertures. Profile
plots are normalized to the {}``hole'' in the center of the hBN
hexagon (value 0) and the nitrogen site (value -1). The light-blue
area indicates as reference an experimental error of 5\% contrast
difference between the atomic sites. \label{fig:BNres}}
\end{figure}

Based on the accurate electrostatic potentials from the all-electron
DFT calculation, we briefly discuss the requirements for identifying
the B and N sublattices in hBN single layer TEM images. For simplicity
we show here the projected electrostatic potentials, limited to a
specific resolution by apertures as indicated. In the real experimental
case, the effect of the microscope would have to be modeled on top
of this by an additional Fourier filter. However, at a given spatial
resolution there can not be more information in the final image, than
is already present in the projected potentials.

The analysis is very simple if only the first Bragg reflection (1-100,
corresponding to $2.16\,\textrm{\AA}$) is transferred. In general,
the image of the hexagonal structure in this case is characterized
by three values, the contrast on the three sub-lattices (B site, N
site, hole in the hexagon). Further, the double-layer image can be
used to verify the effects of residual aberrations. The IAM simulation
with this aperture in Fig. \ref{fig:BNres} illustrates that it is
\emph{in principle} possible to have a different contrast for B and
N sites at this resolution, even though the atomic separation is only
$1.45\,\textrm{\AA}$. It has to be considered as a coincidence that
the ionic character of the material leads to a practically symmetric
contrast in the DFT based simulation. Importantly, our experiment
confirms that no distinction between B and N sites is possible at
this resolution, even with very high signal to noise ratio. 

The situation is not changed by including the $(2-1-10)$ reflection
into the image. This reflection can not carry any {}``which-atom''
information already for symmetry reasons. It corresponds to a $1.45\,\textrm{\AA}$
spacing in a hexagonal structure that adds the same contributions
to the B and N sites (the relative numbers in Fig. \ref{fig:BNres}
are slightly different due to the chosen normalization).

Only after including the $1.08\,\textrm{\AA}$ (2-200) reflection,
we expect a clear difference between B and N sites based on the DFT
model. Also, the relative difference between IAM and DFT based simulation
is smaller. This is reasonable, since at higher resolution there is
more contribution from electrons scattered close to the nucleus, where
bonding effects play a smaller role. Importantly, we note that a \emph{correct}
identification of B and N sites requires that the $1.08\,\textrm{\AA}$
reflection (rather than just the $2.16\,\textrm{\AA}$ as expected
from IAM) is included into the image \emph{without any uncontrolled
phase shifts}. Hence, even if the $1.08\,\textrm{\AA}$ beam could
be transferred in an experiment (it corresponds to a scattering angle
of 40mrad at 80kV), it is not obvious how one could separate the effects
of electron optical aberrations from the intrinsic effect of the sample.
As shown below, this is a complicated task already with only the first
reflection, where one can use the bi-layer hBN regions as reference.

\section{Residual aberrations vs. intrinsic contrast}

For the case of hexagonal boron nitride, it is important to verify
and compensate the effects of residual, non-round aberrations. In
thin hBN samples, it is frequently possible to find a {}``reference''
structure for this purpose just next to the single-layer membrane,
which is a region of double-layer hBN. For the bi-layer area, a symmetric
contrast for the two sub-lattices must be present already due to the
symmetry of the material (B is above N in adjacent layers). The asymmetry
in the experimental intensity profiles for the bi-layer region therefore
shows exclusively the effect of residual aberrations. In a simplistic
approximation, one might assume identical imaging conditions for these
different regions of the sample, imaged simultaneously and separated
by a few nm. However, this has never been verified with a sufficient
precision.

\begin{figure}
\includegraphics[width=0.72\linewidth]{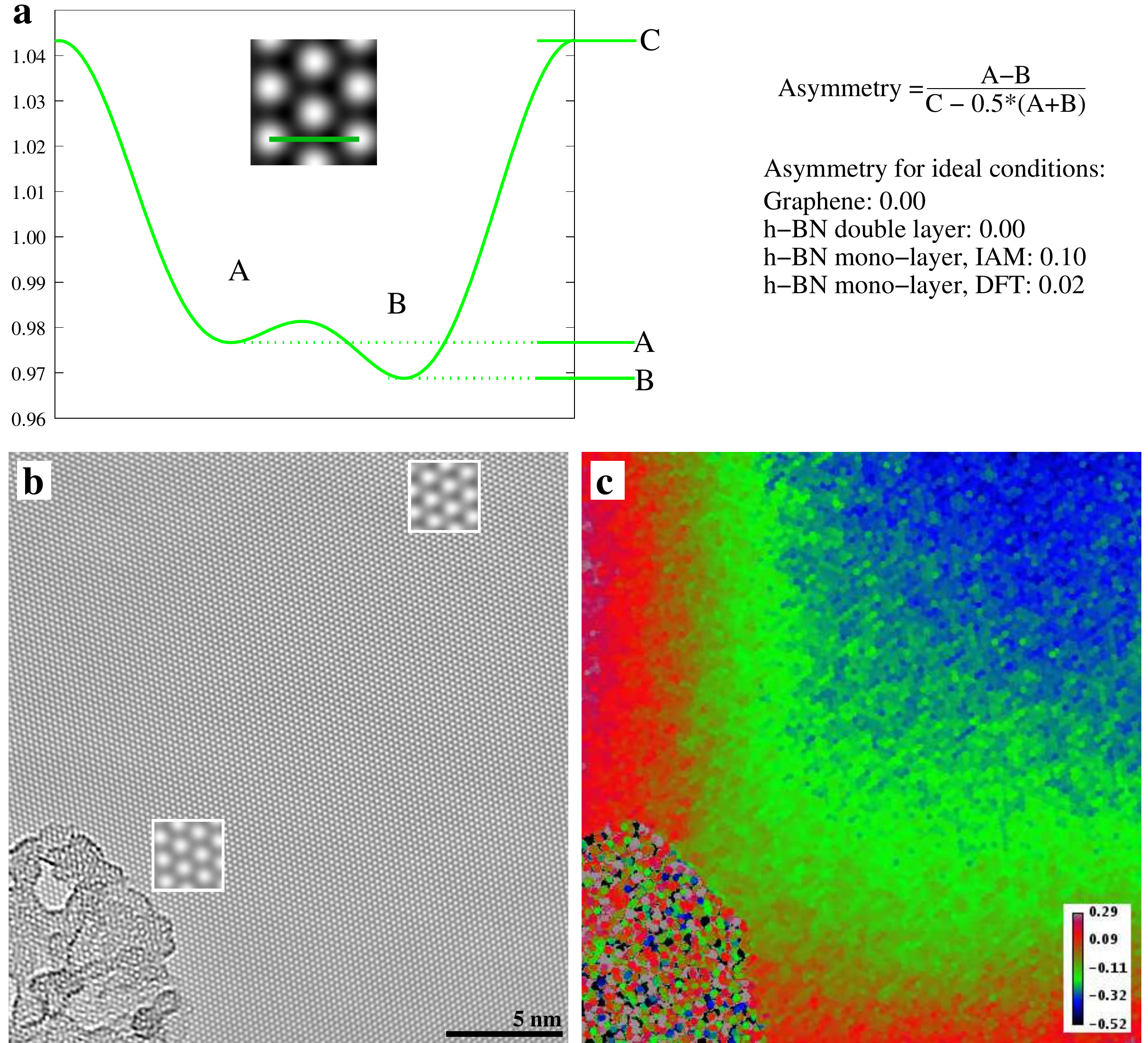}

\caption{Asymmetry definition and measurement. (a) Definition of the {}``Asymmetry''
value, as relative difference of the contrast on the two atomic sites
(A, B), normalized to the total modulation (site C (the hole in the
hexagon), minus the mean value of A andB). Without non-round aberrations,
we expect a zero value for graphene and even-layered hBN, and a non-zero
value for odd-layer hBN. (b) Experimental image of single layer graphene,
recorded with the same conditions as our hBN data, and with similar
signal to noise ratio. Insets show the clearly different images at
separated positions. (c) Automated analysis of the asymmetry value
for every unit cell, revealing the clear variation of this number
across the image.\label{fig:AsymDefAndGr}}
\end{figure}

Fig. \ref{fig:AsymDefAndGr} shows a quantitative analysis of this
point. We define an {}``Asymmetry'' value as the relative contrast
difference between the two atomic sites, normalized to the total modulation.
Under ideal imaging conditions, a symmetric profile (Asymmetry of
zero) would be expected for double-layer (or even-layered) hexagonal
boron nitride regions, as well as for single-layer graphene. For single-layer
hBN, one would expect an asymmetry of 0.10 from the independent-atom
model, and 0.024 from the DFT based result. For triple-layer hBN,
this relative asymmetry is expected as 0.033 from IAM, and 0.008 from
DFT. Hence, it is also possible to discern between IAM and DFT from
a single-layer - triple-layer comparison. We have implemented an automated
method to measure this asymmetry for every unit cell (available as
plug-in for the ImageJ software). The algorithm first locates all
the intensity maxima, which represent the centers of the graphene
or hBN hexagons. Then, it measures the intensity at these maxima (value
C), as well as the intensity on the atomic sites A and B (at a given
distance and orientation relative to the hexagon centers). As shown
in Fig. \ref{fig:AsymDefAndGr}, the asymmetry value does vary considerably
over the field of view, as can be easily verified using images from
large clean areas of single layer graphene.

\begin{figure}
\includegraphics[width=1\linewidth]{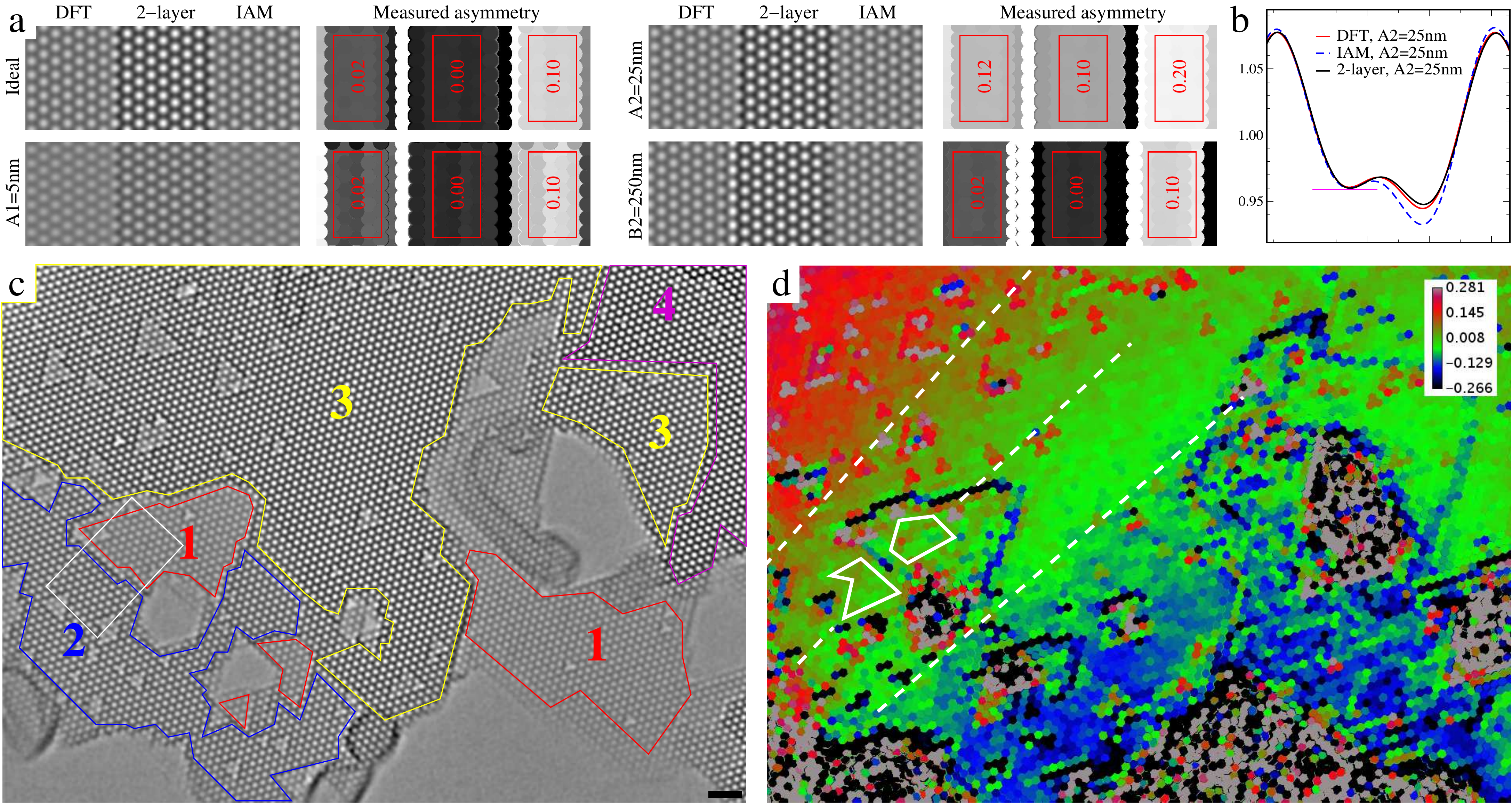}

\caption{(a+b) Simulations for hBN in presence of non-round aberrations. (a)
Simulated images and measured asymmetry for single-layer hBN using
DFT based potentials, bi-layer hBN (same in DFT and IAM), and single-layer
hBN based on IAM. {}``Ideal'' corresponds to Cs=15\ensuremath{µ}m,
defocus $f_{1}=-9\textrm{nm}$, focus spread 4nm. Other images contain
the indicated non-round aberration in addition. (b) Intensity profiles
for the example of A2=25nm. The bi-layer profile was rescaled by a
factor of 1/2, and intensity on the left sub-lattice was adjusted
to the same position (pink line, as in Fig. 5 of the main article).
(c) Larger-scale view experimental image. The area indicated by the
white box was shown and analyzed in Fig. 5 of the main article. Scale
bar is 1nm. (d) Asymmetry analysis of the image. Solid white lines
indicate the areas that were used for the single-layer/bi-layer comparison.
White dashed lines indicate approximately constant asymmetry, separated
by the IAM difference of 0.1. \label{fig:hBNAsymSim}}
\end{figure}

Now, we verify how the asymmetry for hBN depends on residual electron
optical aberrations. We have tested all parameters up to third order
aberrations. As examples, Fig. \ref{fig:hBNAsymSim}a shows simulated
images for hBN in the presence of astigmatism (A1), two-fold astigmatism
(A2), and coma (B2). As expected, the asymmetry depends on some of
these values. However, the \emph{difference} in the asymmetry from
single-layer to bi-layer hBN is not sensitive to these effects. Thus,
one would always expect a sharp step in the asymmetry value when going
from (n) to (n+1) layers in hBN according to IAM, and almost no difference
in the DFT model, independent of small residual aberrations.

Fig. \ref{fig:hBNAsymSim}b shows the intensity profiles for the example
with a two-fold astigmatism of 25nm. Here, the bi-layer profile was
rescaled by a factor of 1/2, and the left sublattice was shifted to
the same position. We then compare the contrast on the other sublattice
(as in the experimental analysis, Fig. 5 of the main article). We
find that contrast difference in this comparison is the same as under
ideal imaging conditions, independent of small non-round aberrations.
Hence, the exact agreement between single-layer and (rescaled) bi-layer
hBN, as found in the experiment, confirms the DFT model and rules
out the IAM, independent of small residual aberrations.

\begin{figure}
\includegraphics[width=0.64\linewidth]{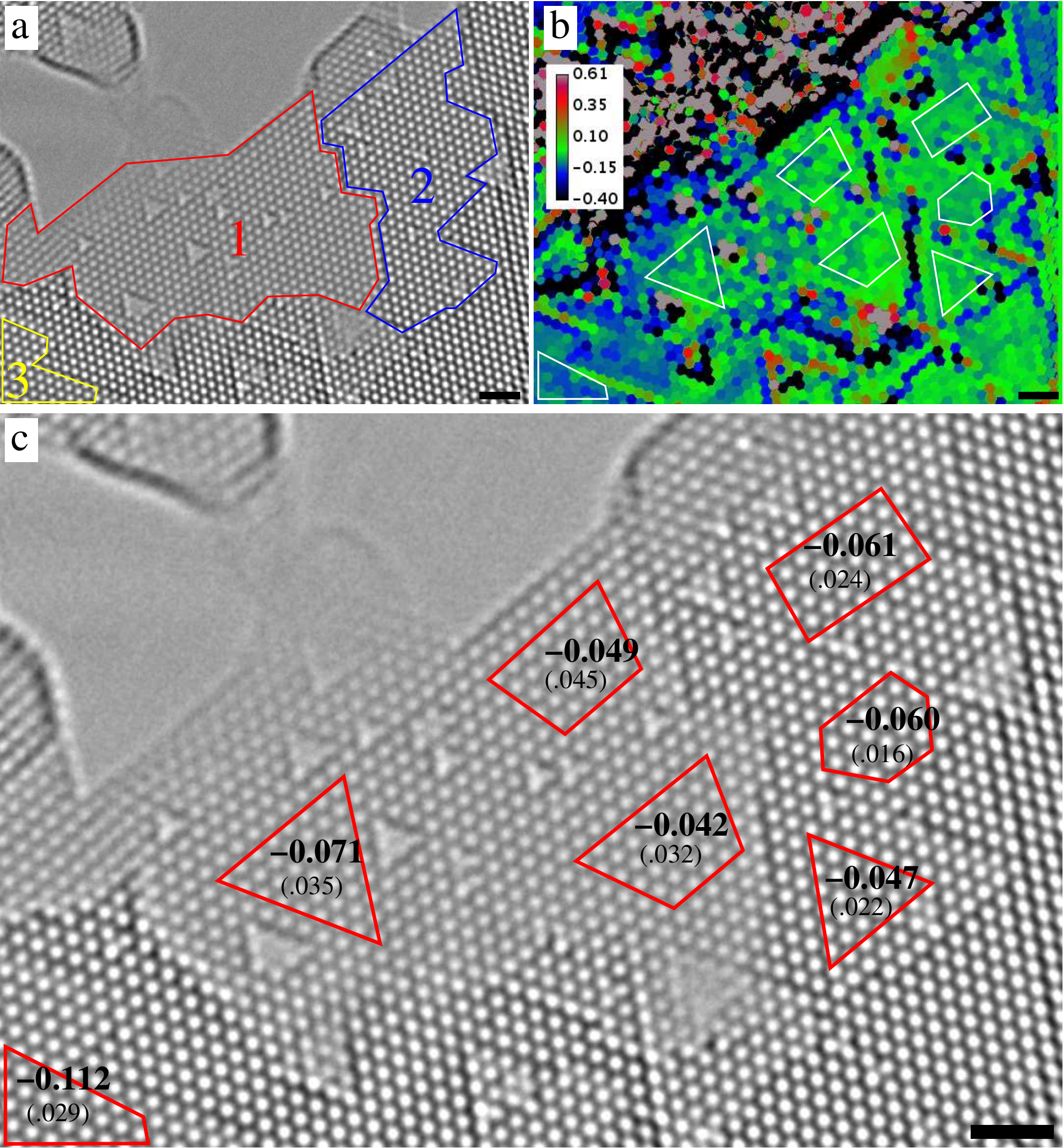}

\caption{Comparison of single- and bi-layer hBN based on the asymmetry measurement.
(a) Experimental image, with single- and bi-layer area indicated.
(b) Asymmetry measurement for every unit cell. Selected regions for
the measurement, away from defects and step edges, are indicated.
(c) Experimental image with measured asymmetry values and regions
indicated. Given in parentheses is the standard deviation of the unit-cell
asymmetry values. Note that the standard deviation is the statistical
uncertainty of the individual unit cell measurement. The uncertainty
of the mean values is better by a factor of at least 5 (for the smallest
region with 25 unit cells). The single- and bi-layer images in this
measurement differ in asymmetry by less than 0.03, which is also the
variation between closely spaced areas of the same thickness. Hence,
they are identical within experimental errors while the 0.1 difference
from IAM can be ruled out. \label{fig:bn100}}
\end{figure}

Fig. \ref{fig:hBNAsymSim}c shows the image of a larger area of ultra-thin
hBN, with the image section that was used in the main article indicated
by a white box. Fig. \ref{fig:hBNAsymSim}d shows the asymmetry measurement
on the entire area. One can see a smooth trend from the top left to
the lower right corner of the image, within large regions of constant
thickness, which must be the effects of varying imaging conditions.
The asymmetry value is unreliable at point defects, step edges, and
in the vacuum region, which however are clearly discerned in the original
image. The single- and bi-layer regions that were analyzed in the
main article are again indicated (solid white lines). Based on the
large-scale trend, these regions were chosen in a way that differences
from residual aberrations are minimized. The white dashed lines indicate
contours of approximately constant asymmetry, with the asymmetry values
differing by 0.1. This corresponds to the IAM difference. As shown
in Fig. \ref{fig:hBNAsymSim}b, a change in two-fold astigmatism of
25nm may produce this amount of asymmetry. However, the assignment
to a specific aberration coefficient is not unique, and also not needed
for the present analysis. In any case, one can easily obtain this
10\% difference as an artefact if the variation of imaging conditions
is not quantitatively verified, already with single-layer and bi-layer
reference region separated by only a few nanometers. 

\begin{figure}
\includegraphics[width=0.99\linewidth]{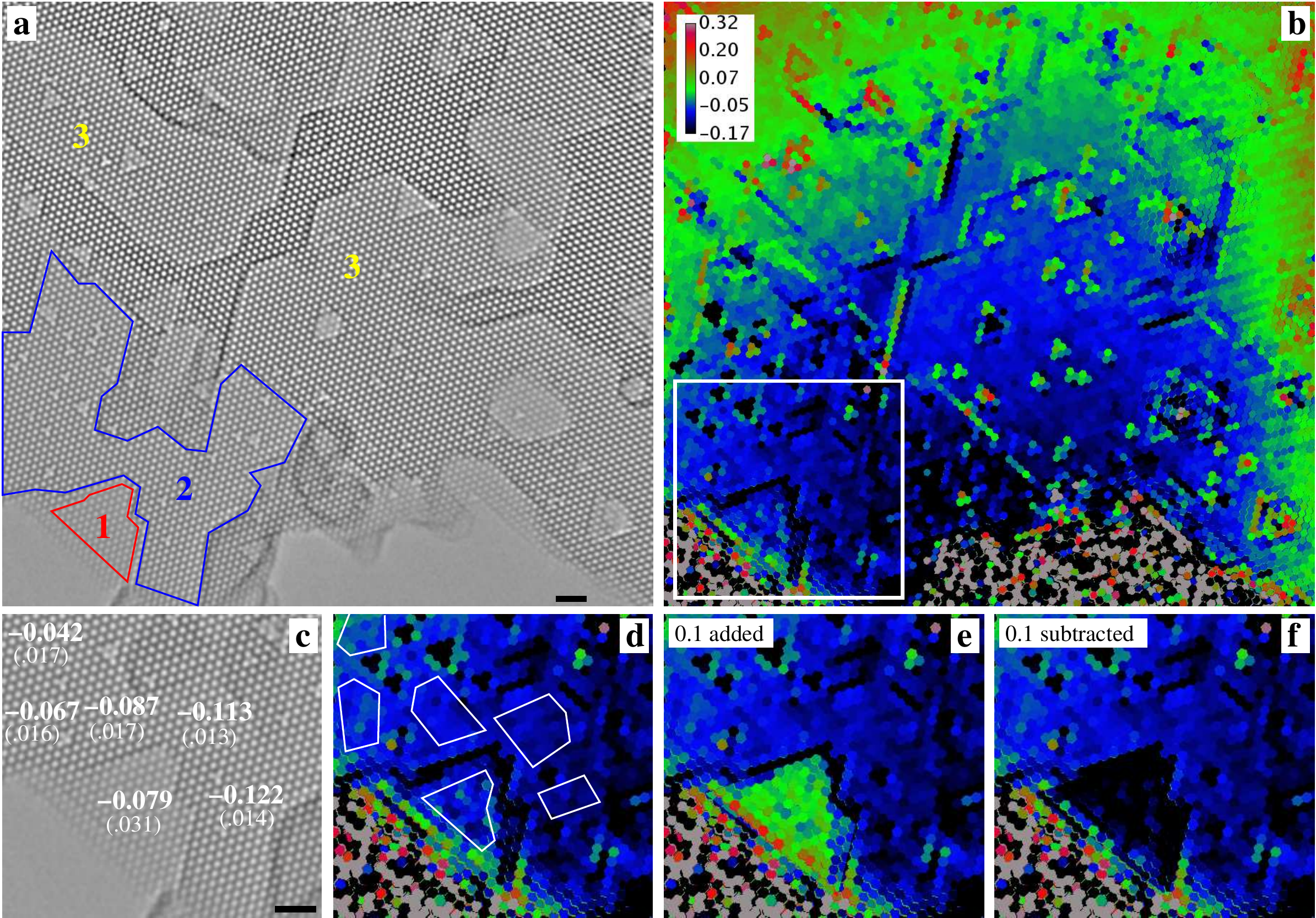}

\caption{Large-area hBN image, including a single-layer region (red) that
is almost surrounded by a double-layer reference (blue). (a) HRTEM
image (average of 30 exposures). The red and blue indicated regions
did not change during the underlying image sequence. (b) Asymmetry
measurement, showing the large scale variations of imaging conditions.
White box indicates the region of interest for c-f. (c+d) Asymmetry
measurement results. The mean asymmetry values and standard deviations
are shown in the image (c), and regions for the corresponding measurements
are shown in (d). All differences between the single-layer area and
surrounding bi-layer region are significantly smaller than expected
from IAM (Again, note that the standard deviation reflects the uncertainty
of the individual measurement; statistical uncertainty of the mean
value is better by a factor of at least 5). (e+f) Asymmetry measurement,
with a value of 0.1 (as expected from IAM) manually added (e) or subtracted
(f) on the single layer region: If the IAM were correct, we should
clearly see such a step between single and double layer areas. Scale
bars are 1nm.\label{fig:IAMaddSubtr}}
\end{figure}

The automated analysis of the hBN images allows to verify our result
for a larger amount of data, which would be very tedious when using
manually selected and averaged unit cells. Fig. \ref{fig:bn100} shows
an example measurement for the comparison of single-layer and bi-layer
hBN using the automated asymmetry analysis. The automated analysis
provides the asymmetry value on every individual unit cell. It is
then easy to extract the mean value and standard deviation (in this
example using 25 or more unit cells in each indicated area). The areas
are chosen such that defects, step edges, and areas where the thickness
has changed during the image sequence are avoided. The identical asymmetry
in single- and bi-layer regions is confirmed. This means that the
images are identical except for a constant factor in the total contrast.
Also, several measurements in an area of constant thickness allow
to further verify errors due to the variations in imaging conditions.

A further example is shown in Fig. \ref{fig:IAMaddSubtr}. Fig. \ref{fig:IAMaddSubtr}a+b
show the larger scale image and the asymmetry analysis on the entire
area. In one corner of the image (Fig. \ref{fig:IAMaddSubtr}a), we
find a single-layer area that is almost surrounded by a double layer
that may serve as reference. We can now easily measure the asymmetry
and its standard deviation for the mono-layer and all surrounding
bi-layer areas: Fig. \ref{fig:IAMaddSubtr}c shows the image section
with the result of the asymmetry measurements, while \ref{fig:IAMaddSubtr}d
shows the color-coded asymmetry image along with the chosen areas.
From all measurements, it is clear that there is no intrinsic difference;
the image of mono-layer hBN is exactly identical to double layer hBN
with half the contrast. All differences between separated regions
are exclusively due to the variation of imaging conditions across
the field of view. In this example, a difference of 0.1 is obtained
over a spatial separation of ca. 7nm (the two furthest separated measurements
in the bi-layer region, Fig. \ref{fig:IAMaddSubtr}c, differ by 0.08),
while several single-layer/double-layer comparisons can be made on
a much shorter distance in the same image. 

Indeed, the difference in asymmetry as predicted by IAM simulations
should be detectable at the present noise level already from visual
inspection of the asymmetry images. To demonstrate this, we have manually
added and subtracted 0.1 to the asymmetry value of single layer regions
(Fig. \ref{fig:IAMaddSubtr}e,f): If the IAM prediction were correct,
one would see such steps very clearly, and this is not the case in
any of our measurements. Also, the step between single/triple, triple/double
or triple/quad layer hBN should be well visible: It is only smaller
in numerical value in our \emph{relative} definition of the asymmetry
but in this value also the noise is equally smaller in thicker sample
areas. Again, the automated analysis allows to screen larger amounts
of data, monitor the variation of imaging conditions with time (in
image sequences), and makes possible errors that may arise in the
manual analysis of a singular measurement highly unlikely.
\end{document}